\documentclass[ 
twocolumn,
showpacs,preprintnumbers,
bibnotes,
 amsmath,amssymb,
 aps,
 pra,
superscriptaddress,
longbibliography,
]{revtex4-1}

\usepackage[version=3]{mhchem} 
\usepackage[dvipsnames]{xcolor}
\usepackage{textcomp}
\usepackage{graphicx}
\usepackage{amsmath}
\usepackage{fixmath}
\usepackage{amsfonts}
\usepackage{amssymb}
\usepackage{braket}
\usepackage{subfigure}

\usepackage{tabularx}
\usepackage{tikz}
\usetikzlibrary{calc,intersections,through,hobby}
\usepackage{cleveref}

\usetikzlibrary{decorations.pathmorphing}
\tikzset{snake it/.style={decorate, decoration=snake}}

\newcommand{\IT}[1]{{\color{black}#1}}

\begin{document}

\title{Non-perturbative effects of deep-strong light-matter interaction in a mesoscopic cavity-QED system}

\author{A.~Kudlis}
\affiliation{Abrikosov Center for Theoretical Physics, MIPT, Dolgoprudnyi, Moscow Region 141701, Russia}
\author{D.~Novokreschenov}
\affiliation{%
 Faculty of Physics, ITMO University, St. Petersburg 197101, Russia}%
\author{I.~Iorsh}
\email{i.iorsh@metalab.ifmo.ru}
\affiliation{Abrikosov Center for Theoretical Physics, MIPT, Dolgoprudnyi, Moscow Region 141701, Russia}%
\affiliation{%
 Faculty of Physics, ITMO University, St. Petersburg 197101, Russia}%
\author{I.~V.~Tokatly}
\affiliation{Nano-Bio Spectroscopy Group and European Theoretical Spectroscopy Facility (ETSF), Departamento de Polímeros y Materiales
Avanzados: Física, Química y Tecnología, Universidad del País Vasco, Avenida Tolosa 72, E-20018 San Sebastián, Spain}
\affiliation{IKERBASQUE, Basque Foundation for Science, 48009 Bilbao, Spain}
\affiliation{Donostia International Physics Center (DIPC), E-20018 Donostia-San Sebastián, Spain}
\affiliation{%
 Faculty of Physics, ITMO University, St. Petersburg 197101, Russia}%

\date{\today}

\begin{abstract}
We consider a system comprising two groups of \IT{quantum} dimers placed in a common electromagnetic cavity, \IT{and controlled by selectively applying a static external potential to one of the groups.} 
\IT{We show that in the regime of deep strong coupling to vacuum electromagnetic fluctuations, the emergent photon-assisted interaction between the dimers, leads to a strongly non-linear quantized cross-polarization response of the first, unbiased group of dimers to the potential applied to the second group. The total polarization shows a series of almost ideal steps whose number and position depends on the parity of the numbers of dimers in the groups.}
This non-perturbative effect is a distinctive feature of mesoscopic systems comprising finite number of dimers and disappears in the thermodynamic limit which is commonly used in the desciption of the generalized Dicke models. 
\end{abstract}

\maketitle

\textit{Introduction.} Polaritonic chemistry~\cite{Ebbesen2016,ribeiro2018polariton}, a novel rapidly developing interdisciplinary field explores the methods to modify chemical properties of materials by placing them inside the optical microcavities. Of particular interest is the regime of strong light matter coupling when the characteristic energy of light-matter interaction exceeds the decay rates of the individual excitations leading to the emergence of the hybrid light-matter quasiparticles, polaritons. Due to the photonic component, polaritons preserve spatial coherence at large distances of the order of the resonant cavity wavelength which as has been shown both theoretically~\cite{Herrera2016} and experimentally~\cite{zhong2017energy} leads to the substantial modification of energy transfer and more generally chemical kinetics in cavity embedded materials. Moreover, for stronger light-matter interaction, when the characteristic energy of light-matter coupling becomes comparable to the excitation energy, the system enters the so-called ultrastrong coupling regime~\cite{Kockum2019} characterized by the substantial modification of the ground state of the system by vacuum fluctuations of cavity electromagnetic field. Ultrastrong coupling was predicted to induce various cavity mediated phase transitions such as  superconductivity ~\cite{thomas2019exploring,curtis2019cavity,sentef2018cavity,schlawin2019cavity,li2020manipulating}, ferroelectric phase transitions~\cite{ashida2020quantum}, topological phase transitions~\cite{PhysRevLett.125.257604, wang2019cavity}, as well as substantial modification of the chemical reactions inside the cavity~\cite{martinez2018can}. 
\begin{figure}
    \centering
    \includegraphics[width = 1\linewidth]{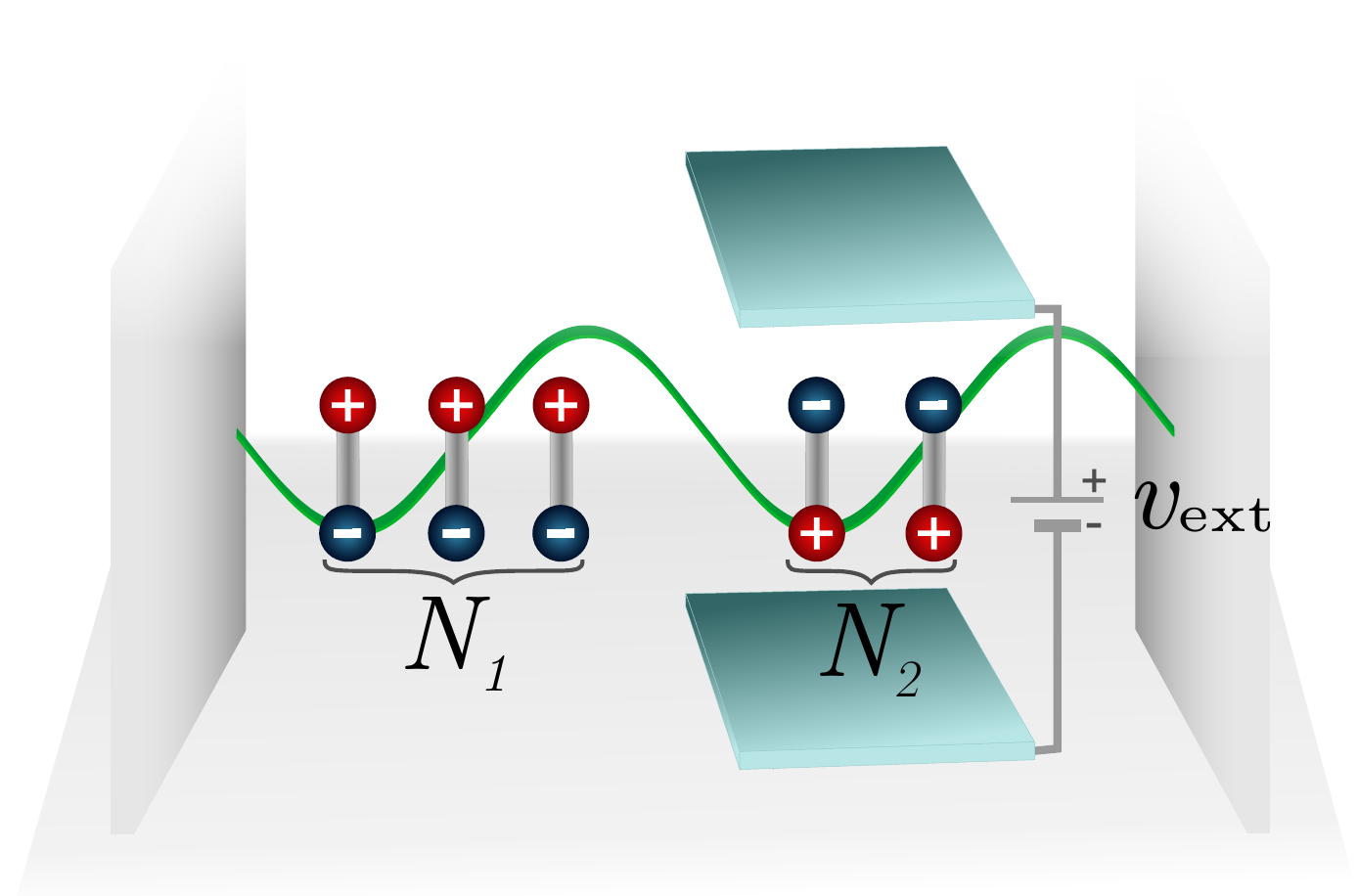}
    \caption{Schematic representation of \IT{the setup: two groups of dimers in the cavity with a polarizing potential applied to one of the groups}. Interaction between dimers is allowed only via photonic mode. In the \IT{shown configuration $N_1=3$ and $N_2=2$, antiferroelectric ordering is observed}.}
    \label{fig:schematic}
\end{figure}

Theoretical description of the ultrastrong coupling between light and matter usually focuses at the two limiting cases. First, one can consider a very small number of two-level systems coupled to the cavity modes of the system. In the limit of a single cavity mode and one two-level system this reduces to the celebrated Rabi model, for which an analytical solution has been found relatively recently~\cite{PhysRevLett.107.100401}. In the opposing limit of large number $N$ of two-level systems, one can exploit the transition to the thermodynamic limit $N\rightarrow \infty$.  It has been shown that in this limit, one may resort to the random phase approximation in the leading order with respect to $1/N$~\cite{Dmytruk2021,Dmytruk2022,Lenk2022}. 

The intermediate case, when the number $N$ is finite but not asymptotically large, corresponding to the mesoscopic regime, is by large terra incognita so far. In this intermediate case, there are not many methods except for the computationally demanding exact diagonalization of the full light-matter Hamiltonian. While, recently new approaches based on the quantum electrodynamics density functional theory (QEDFT) are being developed~\cite{Tokatly2013PRL,Ruggenthaler2014PRA, FarTok2014PRB,Pellegrini2015PRL, Flick2015,ruggenthaler2017groundstate}, their applicability to  generic systems in the ultrastrong coupling regime is still subject of active research.

In this Letter we explore this intermediate regime of finite number of two-level systems in a cavity and demonstrate the emergence of the non-perturbative effects which can not be described within the RPA. Specifically, we consider a system schematically depicted in Fig. 1: $N=N_1+N_2$ dimers are placed in a single mode cavity. Only $N_2$ dimers are subject to the external static potential $v_{ext}$, polarizing the dimers. It is assumed that dimers do not interact directly and are coupled only via the interaction with a cavity electromagnetic mode. 

The Hamiltonian of the system reads
\begin{multline}\label{eqn:starting_ham}
\hat{H}=-T \displaystyle\sum_{i=1}^{N}\hat{\sigma}_{i,x} +v_{\textup{ext}}\displaystyle\sum_{i=N_1+1}^{N}\hat{\sigma}_{i,z}\\
+\dfrac{\hat{p}^2}{2}+\dfrac{\lambda^2}{2}\left(\dfrac{\omega\hat{q}}{\lambda}-\displaystyle\sum_{i=1}^{N}\hat{\sigma}_{i,z}\right)^2,
\end{multline}
The first term describes the tunnelling of the electrons between two states in each dimer with $T$ being intradimer hopping amplitude. An experimental realization which corresponds to this model is an ensemble of the diatomic molecules, or double quantum dots. The tunnelling term thus describes electron hopping between two atoms or two quantum dots. We neglect direct coupling between dimers which is justified since the tunnelling coefficient decays exponentially with the distance. The second term describes the static gating of the one group of the dimers. Indeed, when a static electric field is applied along the dimer direction, the energies of the two dimer sites are split.  The last two terms \IT{correspond to the energy}
$1/8\pi\int(\hat{\textbf{B}}$$^2+\hat{\textbf{E}}$$^2)d\textbf{r}$ of the transverse cavity mode. The magnetic field $\hat{B}=\sqrt{4\pi}\hat{p}$ is proportional to the photon canonical momenta $\hat{p}$. The electric field $\hat{E}=\sqrt{4\pi}(\omega\hat{q}-\lambda \hat{S}_z)$ is related to canonical coordinate $\hat{q}$, \IT{proportional to the electric displacement,} and \IT{$\lambda\sum_{i}\hat{\sigma}_{iz}$ is the total polarization of the system of dimers} with $\lambda$ being the effective light-matter interaction. In what follows we normalize the energy to the cavity photon energy $\omega$. It should be noted the Hamiltonian ~\eqref{eqn:starting_ham} \IT{belongs to} a class of so called generalized Dicke models which have been studied recently~\cite{PhysRevA.94.033850,pilar2020thermodynamics,PhysRevA.97.043820, 10.21468/SciPostPhys.9.5.066,lamata2017digital, shapiro2020universal,PhysRevA.100.022513, akbari2023generalized}. Specifically, it has been shown that the structure of the ground state~\cite{10.21468/SciPostPhys.9.5.066} and thermodynamic properties~\cite{pilar2020thermodynamics} of such systems can substantially deviate from the predictions of the conventional Dicke model.
\begin{figure}[!h]
\centering
\includegraphics[width = 1\linewidth]{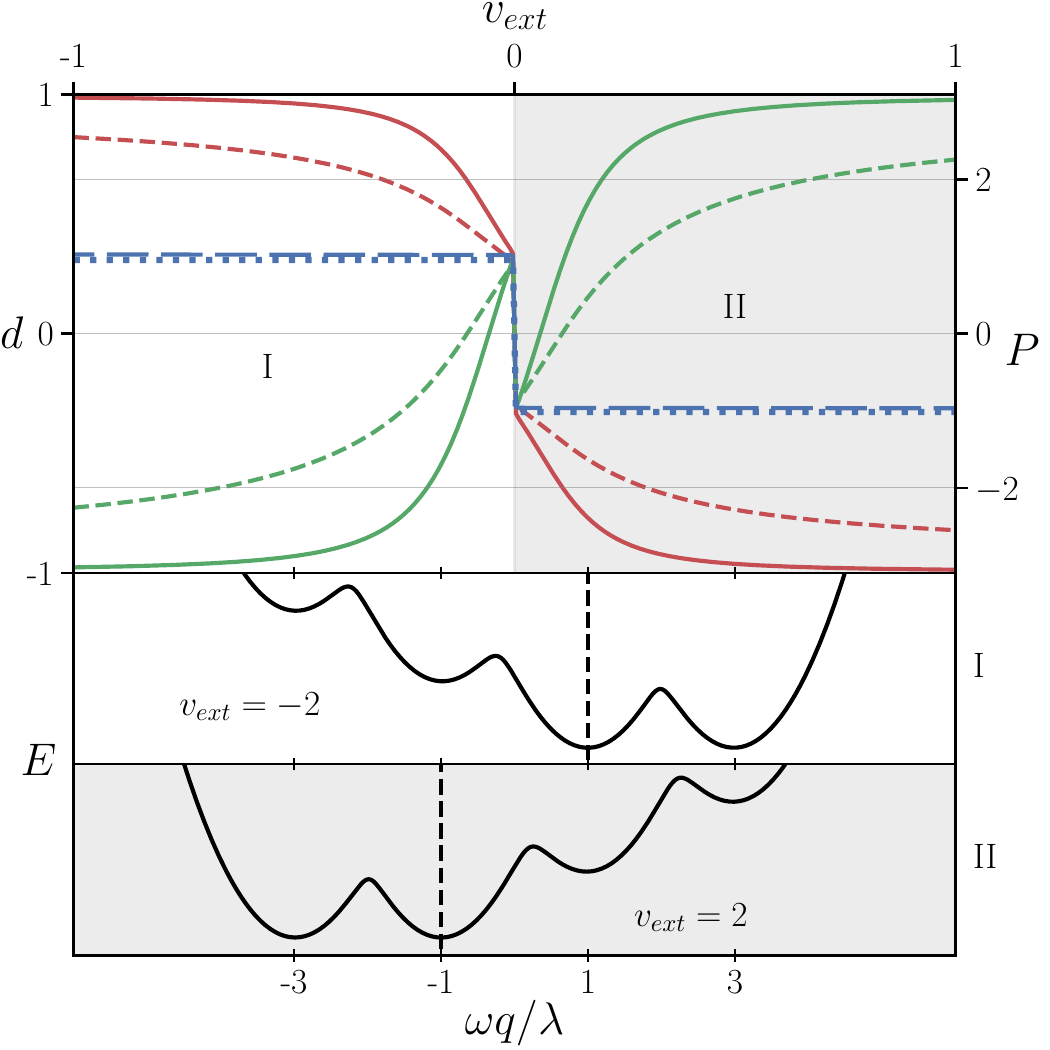}
\includegraphics[width = 1\linewidth]{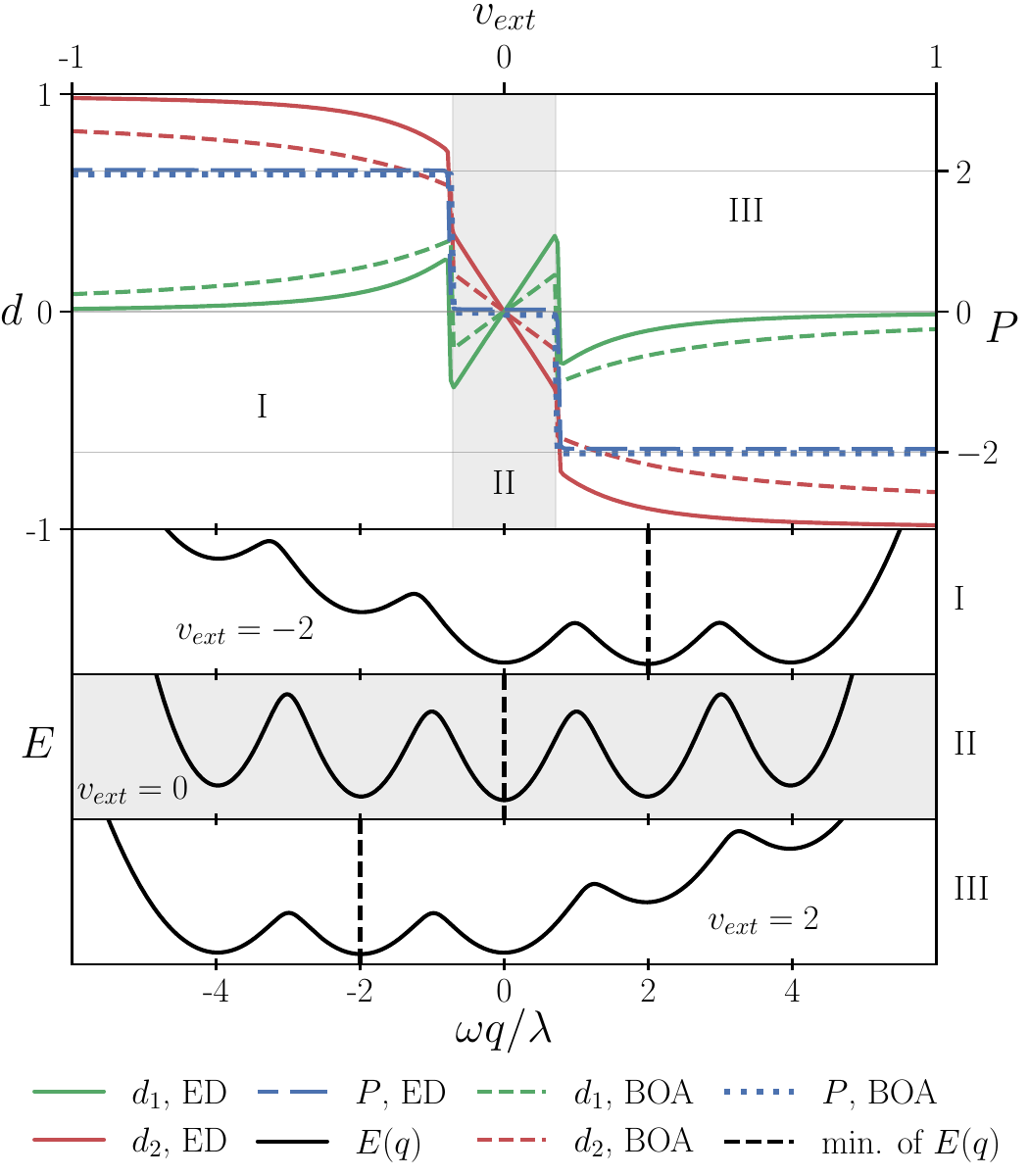}
\caption{Dependencies of $d_1$, $d_2$, and $P$ (upper panel in each figure) on the external field $v_\text{ext}$ \IT{applied to the} second group of dimers, and \IT{semiclassical Born-Oppenheimer (BO) ground state energies as functions of the rescaled photon coordinate $\omega q/\lambda$ in different ranges of $v_{\rm ext}$ (lower panels in each figure)}. The upper figure corresponds to the combination $N_1 = 1$ and $N_2 = 2$. For the lower figure, $N_1 = 2$ and $N_2 = 2$. In both cases, the system parameters: $\lambda = 3$, $\omega = 1$, $T = 1$. BOA and ED denote Born-Oppenheimer approximation and exact diagonalization respectively.}
\label{fig:3+3_3+2_full}
\end{figure}
A common feature of these models are the emergent long-range interactions between the two-level systems facilitated by the exchange of the cavity photon. There was also a certain ambiguity related to the question whether these systems may support a so-called  Dicke superradiant phase transition with the emergence of polarization in the ground state. It is however, now acknowledged that in the gauge invariant formulations of these models, this phase transition is absent in the case of spatially uniform cavity mode profiles~\cite{de2018breakdown,PhysRevB.100.121109,PhysRevB.102.125137}. Specifically, Hamiltonian~\eqref{eqn:starting_ham} is gauge equivalent to a collection of dimers with intradimer hoppings dressed with electromagnetic vector potential via Peierls substitution.

We are interested in the dependence of polarization (which is given by operator $\sigma_z$ for each dimer) for the first group of dimers on the external potential $v_{\rm ext}$ applied to the second group. We used the exact diagonalization to find the ground state of the system.
In what follows we will use the operators of the polarization of the groups of the dimers:$\hat{\mathbf{S}}_{(1,2)}=\sum_{i=1}^{N_{(1,2)}}\hat{\boldsymbol{\sigma}}^{(1,2),i}$ and the total polarization $\hat{\mathbf{S}}=\hat{\mathbf{S}}_1+\hat{\mathbf{S}}_2$

\IT{Figure~\ref{fig:3+3_3+2_full} shows two examples of the dependence on $v_{ext}$ of the average polarization for each group} of dimers $d_{(1,2)}=1/(N_{(1,2)})\sum_{i=1}^{N_{(1,2)}}\langle \sigma_{z}^{(1,2),i} \rangle$, \IT{and of the total polarization $P=\langle S_z\rangle$}. Specifically, we present two cases: $(N_1=1,N_2=2)$ and $N_1=2,N_2=2$. \IT{The main common feature of the presented dependencies are (i) a strongly non-monotonic discontinuous average polarization of first group $d_1(v_{\rm ext})$, and (ii) a sharp, step-like total polarization $P(v_{\rm ext})$.}  We also observe \IT{remarkable differences in the behavior} for these two cases, both at large, and at small $v_{ext}$. Firstly, at large $|v_{ext}|$ the polarizations of the two groups of dimers have opposite signs for $N_1=1$ and the same sign for $N_1=2$. Moreover, at $v_{ext}\approx 0$ the polarization \IT{is almost constant} for $N_1=2$ and has a steep step for $N_1=1$. The plots for other combinations of $(N_1,N_2)$ can be found in the \IT{Supplemental Material}. 

\IT{Importantly, the observed peculiar quantized response} emerges only in the deep strong coupling regime when the dimensionless light-matter coupling strength $\lambda^2/\omega\ge 1$. In this regime, one can neglect the kinetic energy of the harmonic oscillator in Hamiltonian Eq.~\eqref{eqn:starting_ham} and treat $q$ as a classical variable, \IT{which can be viewed as a cavity Born-Oppenheimer approximation  (BOA)~\cite{Flick2017b}}. In this approximation, Hamiltonian reduces to a square  matrix of dimension $(N_1+1)\times (N_2+1)$  and we can find its ground state by finding \IT{the lowest eigenvalue $E(q)$ at each $q$ (the BO surface), and then identifying its global minimum}.  As can be seen in Fig.~\ref{fig:3+3_3+2_full} the polarization found in this approximation (shown with blue dotted lines) \IT{reproduces all main features} of the exact diagonalization (shown by dashed blue lines). 

\begin{figure}[t!]
\centering
\includegraphics[width = 1\linewidth]{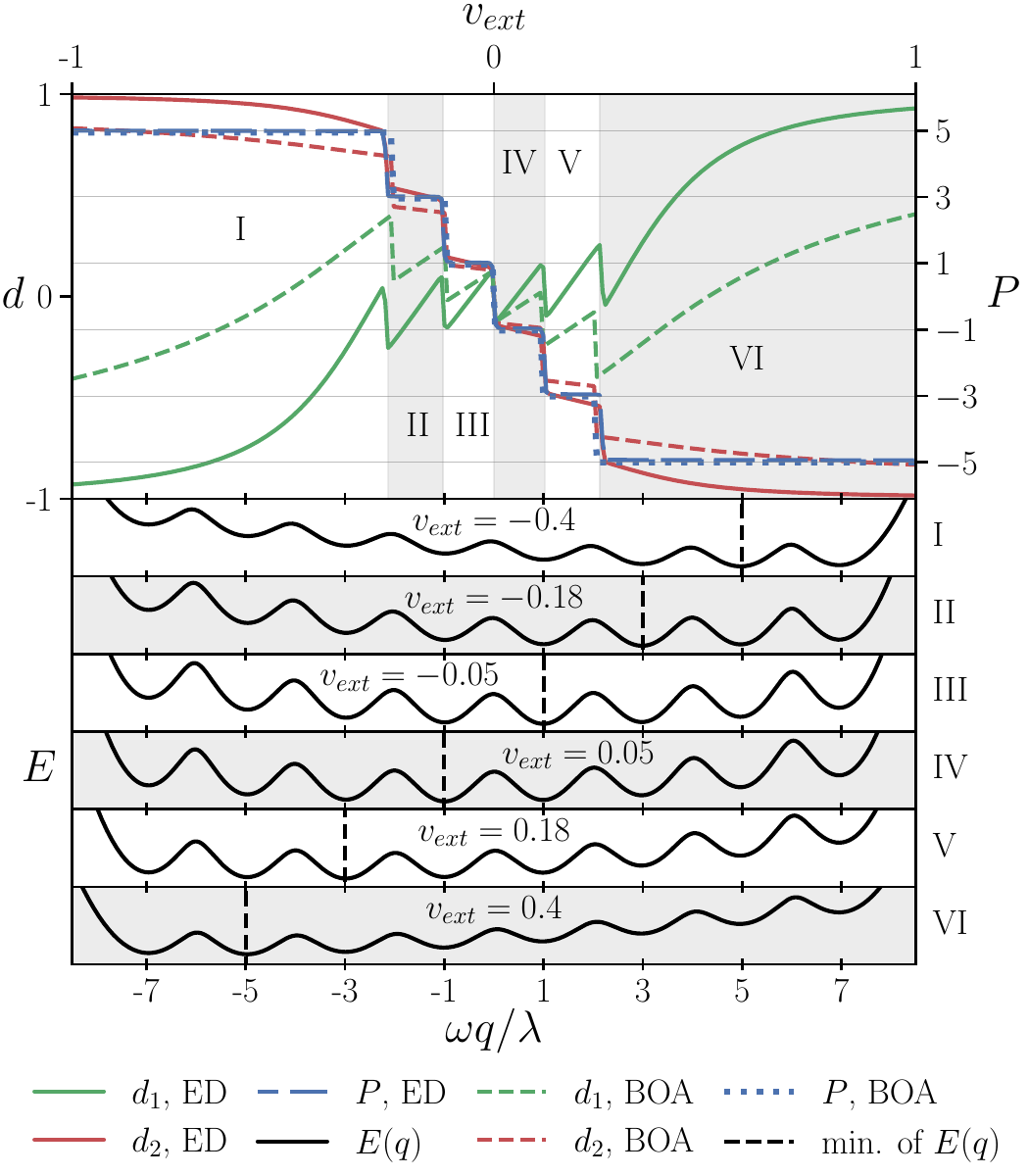}
\caption{\IT{Dependencies of $d_1$, $d_2$, and $P$ on $v_\text{ext}$ applied to the second group of dimers (upper panel), and the BO energies in different ranges of $v_\text{ext}$ as functions of the photon coordinate $q$ (lower panels).} The figure corresponds $N_1 = 1, N_2 = 6$. System parameters: $\lambda = 3$, $\omega = 1$, $T = 1$.}
\label{fig:6+1_full}
\end{figure}


We then make yet another approximation: we first switch off the intra-dimer hopping $T=0$ and then switch it on adiabatically. In the limit $T=0$ the Hamiltonian can be diagonalized exactly. The eigenstates are just the direct products of the eigenstates of $S_{1,z},S_{2,z}$, $|m_1,m_2\rangle =|m_1\rangle\otimes|m_2\rangle$. For each group of dimers there are $N_i+1$ distinct eigenvalues values $m_i=-N_i,-N_i+2\ldots N_i$. The ground state energy is then given by:
\begin{align}\label{eqn:phen_energy}
E_{T=0}= \min_q \left[v_{\text{ext}}m_2 + \dfrac{\lambda^2}{2} \left(\frac{\omega q}{\lambda} - m_1 - m_2\right)^2\right],
\end{align}
In this case $E_{T=0}$ has  local minima at points $\omega q/\lambda = m_1 + m_2 =-N,\dots,N$, in total there are $N+1$ minima. For even $N$(odd number of minima), all minima are located at \IT{even} integer values of $\omega q/\lambda$ $(\dots,-4,-2,0,2,4,\dots)$, and there is a distinguished central minimum with $m=0$. For odd $N$ (even number of minima) they are also at integer points $(\dots,-3,-1,1,3,\dots)$ but \IT{the integers are odd, and} the minimum at $q=0$ is absent.
\IT{At $T=0$ and $v_\text{ext}=0$ all $N+1$ minima in the BO surface are degenerate in energy. At finite $v_\text{ext}$ the number of degenerate valleys reduces to $N_1+1$, while the energies of the remaining $N_2$ minima acquire a linear dependence on $m_2$. This multi-valley structure of the BO surface controlled by $v_\text{ext}$ is the root of the step-like behavior of the total polarization $P$.}

\IT{The degeneracy of different minima is lifted by turning on the intra-dimer hopping $T$ which introduces coupling between states corresponding to different eigenstates of $S_z$ operator at the same $q$.}
The operator $TS_x$ couples the states with $S_z$ projections which differ by $\pm 2$. 
\IT{Therefore} the correction to the energy starts from the second order and \IT{generically lowers the} ground state energy. Moreover, the correction due to the coupling between states with projections different by $2n$ (where $n$ is a positive integer) will be proportional to \textcolor{black}{$(T/\lambda^2)^{n}$}. Apparently for $v_{ext}=0$, the lowest energy corresponds to \IT{minima with smallest} $|q|$ ($q=0$ for even $N$ and $\omega q/\lambda=\pm 1$ for odd $N$). Indeed, \IT{for even $N$, the central minimum at $q=0$ acquires a downward shift that is by an amount \textcolor{black}{$\sim(T/\lambda^2)^{\frac{N}{2}}$} larger compared to the shift of the neighbouring minima at $\omega q/\lambda=\pm 2$.} 
For odd $N$, the two degenerate \IT{central minima at} $\omega q/\lambda = \pm 1$ \IT{are} red shifted with respect to the \IT{closest minima at  $\omega q/\lambda = \pm 3$ by an term} \textcolor{black}{$\sim(T/\lambda^2)^{\frac{N-1}{2}}$}. This simple analysis is confirmed by \IT{computing the BO energies numerically, see} lower panels in Figs.~\ref{fig:3+3_3+2_full} and \ref{fig:6+1_full}. 

\IT{In the case of odd $N$, a week external potential $v_{ext}$ lifts} the degeneracy between the $\omega q/\lambda =\pm 1$ states, and the system falls to one of these minima depending on the sign of $v_{ext}$. This results in the step-like behaviour of polarization at $v_{ext}\approx 0$ shown in Fig.~\ref{fig:3+3_3+2_full} \IT{for $N=3$, and in Fig.~\ref{fig:6+1_full} for $N=7$}.  For \IT{even} $N$ there is a single minimum at $q=0$ and the system remains in this minimum for small $v_{ext}$, 
\IT{as we can see in Fig.~\ref{fig:3+3_3+2_full} for $N=4$}. \IT{In general, a finite $v_{ext}$ favors the extreme values of $m_2=\pm N_2$ in order to minimize term $v_{ext}m_2$. In the limit $T=0$ there are thus $N_1+1$ degenerate minima corresponding to $m_2=-N_2$ (for positive $v_{ext}$)} and for $\omega q/\lambda= m_1+N_2$ and $N_2$ states with values of $m_2$ from $-N_2+2$ to $N_2$. \IT{Nonzero $T$ couples}  states with different $S_z$ \IT{and lifts the degeneracy}. However, in the presence of $v_{ext}$ the \IT{global} minimum does not \IT{always} correspond to the state with smallest $S_z$, because there the energy asymmetry for the states with $S_z$ differing by $\pm 2$. As a result for small $v_{ext}$, the \IT{global energy minimum still corresponds to the valley} with the minimal $S_z$, but as $v_{ext}$ becomes comparable to $T^2/\lambda^2$ the system switches to the state with another value of $S_z$. This results in the stepwise dependence of the polarization on $v_{ext}$ with the width of the steps proportional to $T^2/\lambda^2$. The total number of steps $N_s$ obeys a simple formula
\begin{align}
    N_s=
\begin{cases}
N_2-1, \quad \text{if $N_1$ is odd}\\
N_2, \quad \text{if $N_1$ is even}
\end{cases}
\end{align}
\IT{The validity of this formula is clearly demonstrated by Figs.~\ref{fig:3+3_3+2_full} and \ref{fig:6+1_full}. We provide a chart of characteristic plots for different $N$ in Fig.~\ref{fig:10x10_full} of Supplemental Material.}.

\begin{figure}[t]
     \centering
    \includegraphics[width = 1\linewidth]{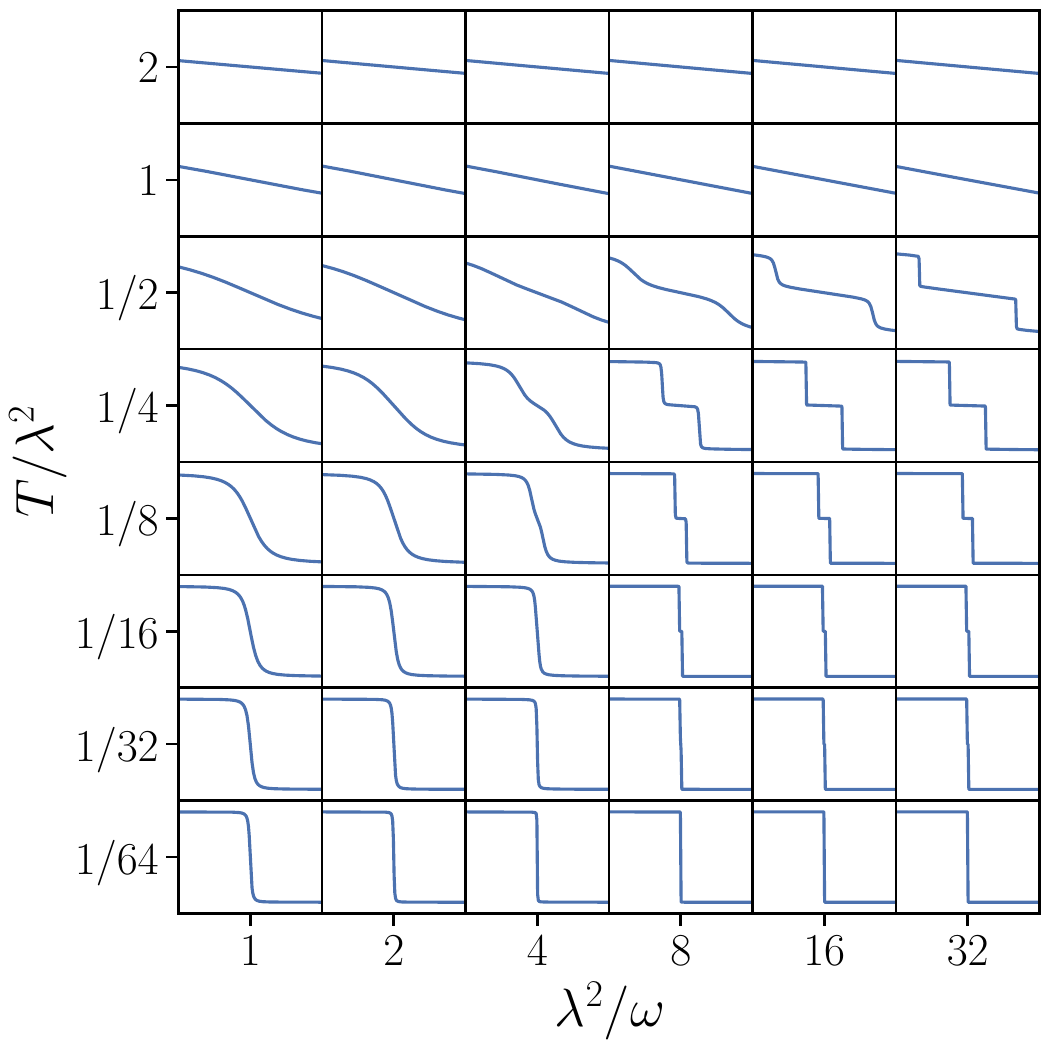}
    \caption{\IT{Polarization $P$ as a function of $v_{ext}$ for different values of $T/\lambda^2$ and $\lambda^2/\omega$. Each cell shows $P$ within limits $[-2.5; 2.5]$ when $v_{ext}$ changes in the range $[-2; 2]$. The system corresponds to $N_1 = 2$ and $N_2 = 2$ at $\lambda=3$.}}
    \label{fig:phase_diag}
\end{figure}

\IT{The quantum nature of electromagnetic field is responsible for transitions between the steps because they occur via tunneling between the corresponding valleys of the BO surface.} To account for this, we replace the variable $q$ by a coherent state $|q\rangle$ of the harmonic oscillator, such that $\langle q| \hat{q} |q \rangle=q$. \IT{The wave function can then be written as a linear combination of the coherent states corresponding to different local minima of the BO surface. The tunneling probability between the minima at different $q_i$ is proportional to the overlap of the corresponding coherent states $\langle q_i|q_j\rangle\sim e^{-(q_i-q_j)^2}\sim e^{-\lambda^2/\omega}$} 
Thus, the parameter $\lambda^2/\omega$ \IT{controls} the coupling between the valley with different $q$, \IT{end $e^{-\lambda^2/\omega}$ determines the width of the steep transition between the states with corresponding $S_z$.} 
To illustrate the dependence of the shape of the steps on $T/\lambda^2$ and ${\lambda^2/\omega}$, in Fig.~\ref{fig:phase_diag} we plot a collection of the step shapes for different values of these two parameters.

\begin{figure}[t!]
     \centering
    \includegraphics[width = 1.\linewidth]{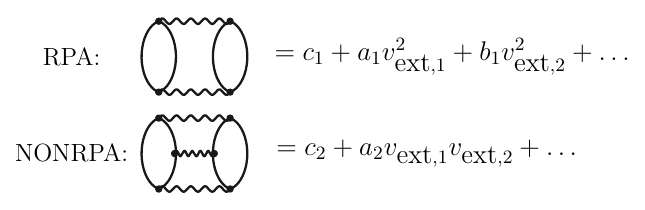}
    \caption{Difference in behavior between the two types of diagrams. It can be seen that the cross susceptibility ($\chi_{12}$), which simply is equal to the cross derivative from diagrams above, in the rpa case is zero if the field on one of the dimers is always zero. Explicit expressions for $a_1,c_1,b_1,c_2,a_2$ are presented in the Supplementary Information.}
    \label{fig:diag}
\end{figure}

It should be emphasized, that the observed step-like behaviour disappears in the thermodynamic limit. \IT{We demonstrate it explicitly using a numerical calculation, the results of which are shown in Fig.~\ref{fig:10x10_full} in SM.} This limit corresponds to $N\rightarrow\infty$ and the scaling of the light-matter interaction as $\lambda\rightarrow \lambda/\sqrt{N}$ \IT{as it is inversely proportional} to the square root of the cavity mode volume. It is now anticipated that the thermodynamic limit of the Dicke and related models can be analyzed within the $1/N$ expansion~\cite{Dmytruk2021,Dmytruk2022,Lenk2022} and the leading order correction \IT{is given by} the RPA-like bubble diagrams. \IT{The second order RPA energy diagram}
is shown in the upper panel of Fig.~\ref{fig:diag}. The bubbles correspond to the dimer \IT{excitation} propagator and wavy lines to the cavity photon propagators. Each bubble has a factor of $N_{1,2}$, \IT{depending on the group of dimers,} and each vertex carries the factor of $\lambda/\sqrt{N}$. Let us \IT{apply} the external potentials $v_{ext,1}$ and $v_{ext,2}$ to \IT{the first and the second} group of dimers, \IT{respectively}. The \IT{differential cross-polarizability of the first group is given by}
$\chi_{12}=\partial^2 E_0/(\partial v_{ext,1}\partial v_{ext,2})|_{v_{ext,1}=0}$. The energy corrections can be expanded with respect to small $v_{ext}$ and calculated explicitly. The result shows, that for the RPA-like diagrams there are no terms in the energy proportional to $v_{ext,1}v_{ext,2}$ and thus \IT{the cross-polarizability in identically zero}. The cross terms appear in the next order of $1/N$ expansion for the diagram shown in the lower panel of Fig.~\ref{fig:diag}. Thus, the magnitude of the cross-polarization scales as $\lambda^6/N$ and vanishes in the thermodynamic limit. Therefore, the effective dimer-dimer interaction emerges only in the case of mesoscopic systems, with finite number of dimers.

An experimental observation of the proposed effect could be realized in the system comprising two spatially separated groups of double quantum dots embedded in a single microwave cavity~\cite{liu2014photon,deng2015coupling,liu2015semiconductor}. It should be noted, that in a realistic system, the cavity photons will have finite lifetime due to the finite cavity quality factor, which may lead to the electroluminiscence in the considered system. While electroluminiscence was previously predicted in the similar set-ups~\cite{PhysRevLett.116.113601}, a self consistent description of the spectral and statistical properties of the emission would require an input-output formalism supplemented with a density matrix Master equation tailored for the ultrastrong coupling regime~\cite{akbari2023generalized}.

To conclude, we have shown that mesoscopic systems in the ultrastrong coupling regime demonstrate the non-perturbative behaviour, not captured neither by the weak coupling perturbation nor by the $1/N$ expansion conventionally used for the description of the Dicke-like Hamiltonians. Specifically, in the system comprising two group of dimers in a common cavity, we have revealed a quantized dependence of the \IT{cross-polarization and the total polarization on the external potential applied selectively to one of the groups}. We give a qualitative explanation to the discovered effect and explain why it occurs only for finite numbers of dimers and deep strong light-matter coupling regime. These results open new routes to exploring physics of deep strong light matter coupling in mesoscopic systems.

\textit{Acknowledgement}.
I.V.T. acknowledges support by Grupos Consolidados UPV/EHU del Gobierno Vasco (Grant No. IT1249-19) and by Spanish MICINN (Project No. PID2020-112811GB-I00).

\appendix

\bibliography{cavityQED.bib}

\begin{thebibliography}{40}%
\makeatletter
\providecommand \@ifxundefined [1]{%
 \@ifx{#1\undefined}
}%
\providecommand \@ifnum [1]{%
 \ifnum #1\expandafter \@firstoftwo
 \else \expandafter \@secondoftwo
 \fi
}%
\providecommand \@ifx [1]{%
 \ifx #1\expandafter \@firstoftwo
 \else \expandafter \@secondoftwo
 \fi
}%
\providecommand \natexlab [1]{#1}%
\providecommand \enquote  [1]{``#1''}%
\providecommand \bibnamefont  [1]{#1}%
\providecommand \bibfnamefont [1]{#1}%
\providecommand \citenamefont [1]{#1}%
\providecommand \href@noop [0]{\@secondoftwo}%
\providecommand \href [0]{\begingroup \@sanitize@url \@href}%
\providecommand \@href[1]{\@@startlink{#1}\@@href}%
\providecommand \@@href[1]{\endgroup#1\@@endlink}%
\providecommand \@sanitize@url [0]{\catcode `\\12\catcode `\$12\catcode
  `\&12\catcode `\#12\catcode `\^12\catcode `\_12\catcode `\%12\relax}%
\providecommand \@@startlink[1]{}%
\providecommand \@@endlink[0]{}%
\providecommand \url  [0]{\begingroup\@sanitize@url \@url }%
\providecommand \@url [1]{\endgroup\@href {#1}{\urlprefix }}%
\providecommand \urlprefix  [0]{URL }%
\providecommand \Eprint [0]{\href }%
\providecommand \doibase [0]{http://dx.doi.org/}%
\providecommand \selectlanguage [0]{\@gobble}%
\providecommand \bibinfo  [0]{\@secondoftwo}%
\providecommand \bibfield  [0]{\@secondoftwo}%
\providecommand \translation [1]{[#1]}%
\providecommand \BibitemOpen [0]{}%
\providecommand \bibitemStop [0]{}%
\providecommand \bibitemNoStop [0]{.\EOS\space}%
\providecommand \EOS [0]{\spacefactor3000\relax}%
\providecommand \BibitemShut  [1]{\csname bibitem#1\endcsname}%
\let\auto@bib@innerbib\@empty
\bibitem [{\citenamefont {Ebbesen}(2016)}]{Ebbesen2016}%
  \BibitemOpen
  \bibfield  {author} {\bibinfo {author} {\bibfnamefont {Thomas~W.}\
  \bibnamefont {Ebbesen}},\ }\bibfield  {title} {\enquote {\bibinfo {title}
  {Hybrid light-matter states in a molecular and material science
  perspective},}\ }\href {\doibase 10.1021/acs.accounts.6b00295} {\bibfield
  {journal} {\bibinfo  {journal} {Accounts of Chemical Research}\ }\textbf
  {\bibinfo {volume} {49}},\ \bibinfo {pages} {2403--2412} (\bibinfo {year}
  {2016})}\BibitemShut {NoStop}%
\bibitem [{\citenamefont {Ribeiro}\ \emph {et~al.}(2018)\citenamefont
  {Ribeiro}, \citenamefont {Mart{\'\i}nez-Mart{\'\i}nez}, \citenamefont {Du},
  \citenamefont {Campos-Gonzalez-Angulo},\ and\ \citenamefont
  {Yuen-Zhou}}]{ribeiro2018polariton}%
  \BibitemOpen
  \bibfield  {author} {\bibinfo {author} {\bibfnamefont {Raphael~F}\
  \bibnamefont {Ribeiro}}, \bibinfo {author} {\bibfnamefont {Luis~A}\
  \bibnamefont {Mart{\'\i}nez-Mart{\'\i}nez}}, \bibinfo {author} {\bibfnamefont
  {Matthew}\ \bibnamefont {Du}}, \bibinfo {author} {\bibfnamefont {Jorge}\
  \bibnamefont {Campos-Gonzalez-Angulo}}, \ and\ \bibinfo {author}
  {\bibfnamefont {Joel}\ \bibnamefont {Yuen-Zhou}},\ }\bibfield  {title}
  {\enquote {\bibinfo {title} {Polariton chemistry: controlling molecular
  dynamics with optical cavities},}\ }\href@noop {} {\bibfield  {journal}
  {\bibinfo  {journal} {Chemical science}\ }\textbf {\bibinfo {volume} {9}},\
  \bibinfo {pages} {6325--6339} (\bibinfo {year} {2018})}\BibitemShut {NoStop}%
\bibitem [{\citenamefont {Herrera}\ and\ \citenamefont
  {Spano}(2016)}]{Herrera2016}%
  \BibitemOpen
  \bibfield  {author} {\bibinfo {author} {\bibfnamefont {Felipe}\ \bibnamefont
  {Herrera}}\ and\ \bibinfo {author} {\bibfnamefont {Frank~C.}\ \bibnamefont
  {Spano}},\ }\bibfield  {title} {\enquote {\bibinfo {title} {Cavity-controlled
  chemistry in molecular ensembles},}\ }\href {\doibase
  10.1103/physrevlett.116.238301} {\bibfield  {journal} {\bibinfo  {journal}
  {Phys. Rev. Lett.}\ }\textbf {\bibinfo {volume} {116}},\ \bibinfo {pages}
  {238301} (\bibinfo {year} {2016})}\BibitemShut {NoStop}%
\bibitem [{\citenamefont {Zhong}\ \emph {et~al.}(2017)\citenamefont {Zhong},
  \citenamefont {Chervy}, \citenamefont {Zhang}, \citenamefont {Thomas},
  \citenamefont {George}, \citenamefont {Genet}, \citenamefont {Hutchison},\
  and\ \citenamefont {Ebbesen}}]{zhong2017energy}%
  \BibitemOpen
  \bibfield  {author} {\bibinfo {author} {\bibfnamefont {Xiaolan}\ \bibnamefont
  {Zhong}}, \bibinfo {author} {\bibfnamefont {Thibault}\ \bibnamefont
  {Chervy}}, \bibinfo {author} {\bibfnamefont {Lei}\ \bibnamefont {Zhang}},
  \bibinfo {author} {\bibfnamefont {Anoop}\ \bibnamefont {Thomas}}, \bibinfo
  {author} {\bibfnamefont {Jino}\ \bibnamefont {George}}, \bibinfo {author}
  {\bibfnamefont {Cyriaque}\ \bibnamefont {Genet}}, \bibinfo {author}
  {\bibfnamefont {James~A}\ \bibnamefont {Hutchison}}, \ and\ \bibinfo {author}
  {\bibfnamefont {Thomas~W}\ \bibnamefont {Ebbesen}},\ }\bibfield  {title}
  {\enquote {\bibinfo {title} {Energy transfer between spatially separated
  entangled molecules},}\ }\href@noop {} {\bibfield  {journal} {\bibinfo
  {journal} {Angewandte Chemie}\ }\textbf {\bibinfo {volume} {129}},\ \bibinfo
  {pages} {9162--9166} (\bibinfo {year} {2017})}\BibitemShut {NoStop}%
\bibitem [{\citenamefont {Kockum}\ \emph {et~al.}(2019)\citenamefont {Kockum},
  \citenamefont {Miranowicz}, \citenamefont {Liberato}, \citenamefont
  {Savasta},\ and\ \citenamefont {Nori}}]{Kockum2019}%
  \BibitemOpen
  \bibfield  {author} {\bibinfo {author} {\bibfnamefont {Anton~Frisk}\
  \bibnamefont {Kockum}}, \bibinfo {author} {\bibfnamefont {Adam}\ \bibnamefont
  {Miranowicz}}, \bibinfo {author} {\bibfnamefont {Simone~De}\ \bibnamefont
  {Liberato}}, \bibinfo {author} {\bibfnamefont {Salvatore}\ \bibnamefont
  {Savasta}}, \ and\ \bibinfo {author} {\bibfnamefont {Franco}\ \bibnamefont
  {Nori}},\ }\bibfield  {title} {\enquote {\bibinfo {title} {Ultrastrong
  coupling between light and matter},}\ }\href {\doibase
  10.1038/s42254-018-0006-2} {\bibfield  {journal} {\bibinfo  {journal} {Nature
  Reviews Physics}\ }\textbf {\bibinfo {volume} {1}},\ \bibinfo {pages}
  {19--40} (\bibinfo {year} {2019})}\BibitemShut {NoStop}%
\bibitem [{\citenamefont {Thomas}\ \emph {et~al.}(2019)\citenamefont {Thomas},
  \citenamefont {Devaux}, \citenamefont {Nagarajan}, \citenamefont {Chervy},
  \citenamefont {Seidel}, \citenamefont {Hagenm{\"u}ller}, \citenamefont
  {Sch{\"u}tz}, \citenamefont {Schachenmayer}, \citenamefont {Genet},
  \citenamefont {Pupillo} \emph {et~al.}}]{thomas2019exploring}%
  \BibitemOpen
  \bibfield  {author} {\bibinfo {author} {\bibfnamefont {Anoop}\ \bibnamefont
  {Thomas}}, \bibinfo {author} {\bibfnamefont {Elo{\"\i}se}\ \bibnamefont
  {Devaux}}, \bibinfo {author} {\bibfnamefont {Kalaivanan}\ \bibnamefont
  {Nagarajan}}, \bibinfo {author} {\bibfnamefont {Thibault}\ \bibnamefont
  {Chervy}}, \bibinfo {author} {\bibfnamefont {Marcus}\ \bibnamefont {Seidel}},
  \bibinfo {author} {\bibfnamefont {David}\ \bibnamefont {Hagenm{\"u}ller}},
  \bibinfo {author} {\bibfnamefont {Stefan}\ \bibnamefont {Sch{\"u}tz}},
  \bibinfo {author} {\bibfnamefont {Johannes}\ \bibnamefont {Schachenmayer}},
  \bibinfo {author} {\bibfnamefont {Cyriaque}\ \bibnamefont {Genet}}, \bibinfo
  {author} {\bibfnamefont {Guido}\ \bibnamefont {Pupillo}},  \emph {et~al.},\
  }\bibfield  {title} {\enquote {\bibinfo {title} {Exploring superconductivity
  under strong coupling with the vacuum electromagnetic field},}\ }\href@noop
  {} {\bibfield  {journal} {\bibinfo  {journal} {arXiv preprint
  arXiv:1911.01459}\ } (\bibinfo {year} {2019})}\BibitemShut {NoStop}%
\bibitem [{\citenamefont {Curtis}\ \emph {et~al.}(2019)\citenamefont {Curtis},
  \citenamefont {Raines}, \citenamefont {Allocca}, \citenamefont {Hafezi},\
  and\ \citenamefont {Galitski}}]{curtis2019cavity}%
  \BibitemOpen
  \bibfield  {author} {\bibinfo {author} {\bibfnamefont {Jonathan~B}\
  \bibnamefont {Curtis}}, \bibinfo {author} {\bibfnamefont {Zachary~M}\
  \bibnamefont {Raines}}, \bibinfo {author} {\bibfnamefont {Andrew~A}\
  \bibnamefont {Allocca}}, \bibinfo {author} {\bibfnamefont {Mohammad}\
  \bibnamefont {Hafezi}}, \ and\ \bibinfo {author} {\bibfnamefont {Victor~M}\
  \bibnamefont {Galitski}},\ }\bibfield  {title} {\enquote {\bibinfo {title}
  {Cavity quantum eliashberg enhancement of superconductivity},}\ }\href@noop
  {} {\bibfield  {journal} {\bibinfo  {journal} {Physical review letters}\
  }\textbf {\bibinfo {volume} {122}},\ \bibinfo {pages} {167002} (\bibinfo
  {year} {2019})}\BibitemShut {NoStop}%
\bibitem [{\citenamefont {Sentef}\ \emph {et~al.}(2018)\citenamefont {Sentef},
  \citenamefont {Ruggenthaler},\ and\ \citenamefont
  {Rubio}}]{sentef2018cavity}%
  \BibitemOpen
  \bibfield  {author} {\bibinfo {author} {\bibfnamefont {Michael~A}\
  \bibnamefont {Sentef}}, \bibinfo {author} {\bibfnamefont {Michael}\
  \bibnamefont {Ruggenthaler}}, \ and\ \bibinfo {author} {\bibfnamefont
  {Angel}\ \bibnamefont {Rubio}},\ }\bibfield  {title} {\enquote {\bibinfo
  {title} {Cavity quantum-electrodynamical polaritonically enhanced
  electron-phonon coupling and its influence on superconductivity},}\
  }\href@noop {} {\bibfield  {journal} {\bibinfo  {journal} {Science advances}\
  }\textbf {\bibinfo {volume} {4}},\ \bibinfo {pages} {eaau6969} (\bibinfo
  {year} {2018})}\BibitemShut {NoStop}%
\bibitem [{\citenamefont {Schlawin}\ \emph {et~al.}(2019)\citenamefont
  {Schlawin}, \citenamefont {Cavalleri},\ and\ \citenamefont
  {Jaksch}}]{schlawin2019cavity}%
  \BibitemOpen
  \bibfield  {author} {\bibinfo {author} {\bibfnamefont {Frank}\ \bibnamefont
  {Schlawin}}, \bibinfo {author} {\bibfnamefont {Andrea}\ \bibnamefont
  {Cavalleri}}, \ and\ \bibinfo {author} {\bibfnamefont {Dieter}\ \bibnamefont
  {Jaksch}},\ }\bibfield  {title} {\enquote {\bibinfo {title} {Cavity-mediated
  electron-photon superconductivity},}\ }\href@noop {} {\bibfield  {journal}
  {\bibinfo  {journal} {Physical review letters}\ }\textbf {\bibinfo {volume}
  {122}},\ \bibinfo {pages} {133602} (\bibinfo {year} {2019})}\BibitemShut
  {NoStop}%
\bibitem [{\citenamefont {Li}\ and\ \citenamefont
  {Eckstein}(2020)}]{li2020manipulating}%
  \BibitemOpen
  \bibfield  {author} {\bibinfo {author} {\bibfnamefont {Jiajun}\ \bibnamefont
  {Li}}\ and\ \bibinfo {author} {\bibfnamefont {Martin}\ \bibnamefont
  {Eckstein}},\ }\bibfield  {title} {\enquote {\bibinfo {title} {Manipulating
  intertwined orders in solids with quantum light},}\ }\href@noop {} {\bibfield
   {journal} {\bibinfo  {journal} {Physical Review Letters}\ }\textbf {\bibinfo
  {volume} {125}},\ \bibinfo {pages} {217402} (\bibinfo {year}
  {2020})}\BibitemShut {NoStop}%
\bibitem [{\citenamefont {Ashida}\ \emph {et~al.}(2020)\citenamefont {Ashida},
  \citenamefont {{\.I}mamo{\u{g}}lu}, \citenamefont {Faist}, \citenamefont
  {Jaksch}, \citenamefont {Cavalleri},\ and\ \citenamefont
  {Demler}}]{ashida2020quantum}%
  \BibitemOpen
  \bibfield  {author} {\bibinfo {author} {\bibfnamefont {Yuto}\ \bibnamefont
  {Ashida}}, \bibinfo {author} {\bibfnamefont {Ata{\c{c}}}\ \bibnamefont
  {{\.I}mamo{\u{g}}lu}}, \bibinfo {author} {\bibfnamefont {J{\'e}r{\^o}me}\
  \bibnamefont {Faist}}, \bibinfo {author} {\bibfnamefont {Dieter}\
  \bibnamefont {Jaksch}}, \bibinfo {author} {\bibfnamefont {Andrea}\
  \bibnamefont {Cavalleri}}, \ and\ \bibinfo {author} {\bibfnamefont {Eugene}\
  \bibnamefont {Demler}},\ }\bibfield  {title} {\enquote {\bibinfo {title}
  {Quantum electrodynamic control of matter: Cavity-enhanced ferroelectric
  phase transition},}\ }\href@noop {} {\bibfield  {journal} {\bibinfo
  {journal} {Physical Review X}\ }\textbf {\bibinfo {volume} {10}},\ \bibinfo
  {pages} {041027} (\bibinfo {year} {2020})}\BibitemShut {NoStop}%
\bibitem [{\citenamefont {Guerci}\ \emph {et~al.}(2020)\citenamefont {Guerci},
  \citenamefont {Simon},\ and\ \citenamefont {Mora}}]{PhysRevLett.125.257604}%
  \BibitemOpen
  \bibfield  {author} {\bibinfo {author} {\bibfnamefont {Daniele}\ \bibnamefont
  {Guerci}}, \bibinfo {author} {\bibfnamefont {Pascal}\ \bibnamefont {Simon}},
  \ and\ \bibinfo {author} {\bibfnamefont {Christophe}\ \bibnamefont {Mora}},\
  }\bibfield  {title} {\enquote {\bibinfo {title} {Superradiant phase
  transition in electronic systems and emergent topological phases},}\ }\href
  {\doibase 10.1103/PhysRevLett.125.257604} {\bibfield  {journal} {\bibinfo
  {journal} {Phys. Rev. Lett.}\ }\textbf {\bibinfo {volume} {125}},\ \bibinfo
  {pages} {257604} (\bibinfo {year} {2020})}\BibitemShut {NoStop}%
\bibitem [{\citenamefont {Wang}\ \emph {et~al.}(2019)\citenamefont {Wang},
  \citenamefont {Ronca},\ and\ \citenamefont {Sentef}}]{wang2019cavity}%
  \BibitemOpen
  \bibfield  {author} {\bibinfo {author} {\bibfnamefont {Xiao}\ \bibnamefont
  {Wang}}, \bibinfo {author} {\bibfnamefont {Enrico}\ \bibnamefont {Ronca}}, \
  and\ \bibinfo {author} {\bibfnamefont {Michael~A}\ \bibnamefont {Sentef}},\
  }\bibfield  {title} {\enquote {\bibinfo {title} {Cavity quantum
  electrodynamical chern insulator: Towards light-induced quantized anomalous
  hall effect in graphene},}\ }\href@noop {} {\bibfield  {journal} {\bibinfo
  {journal} {Physical Review B}\ }\textbf {\bibinfo {volume} {99}},\ \bibinfo
  {pages} {235156} (\bibinfo {year} {2019})}\BibitemShut {NoStop}%
\bibitem [{\citenamefont {Mart{\'\i}nez-Mart{\'\i}nez}\ \emph
  {et~al.}(2018)\citenamefont {Mart{\'\i}nez-Mart{\'\i}nez}, \citenamefont
  {Ribeiro}, \citenamefont {Campos-Gonz{\'a}lez-Angulo},\ and\ \citenamefont
  {Yuen-Zhou}}]{martinez2018can}%
  \BibitemOpen
  \bibfield  {author} {\bibinfo {author} {\bibfnamefont {Luis~A}\ \bibnamefont
  {Mart{\'\i}nez-Mart{\'\i}nez}}, \bibinfo {author} {\bibfnamefont {Raphael~F}\
  \bibnamefont {Ribeiro}}, \bibinfo {author} {\bibfnamefont {Jorge}\
  \bibnamefont {Campos-Gonz{\'a}lez-Angulo}}, \ and\ \bibinfo {author}
  {\bibfnamefont {Joel}\ \bibnamefont {Yuen-Zhou}},\ }\bibfield  {title}
  {\enquote {\bibinfo {title} {Can ultrastrong coupling change ground-state
  chemical reactions?}}\ }\href@noop {} {\bibfield  {journal} {\bibinfo
  {journal} {ACS Photonics}\ }\textbf {\bibinfo {volume} {5}},\ \bibinfo
  {pages} {167--176} (\bibinfo {year} {2018})}\BibitemShut {NoStop}%
\bibitem [{\citenamefont {Braak}(2011)}]{PhysRevLett.107.100401}%
  \BibitemOpen
  \bibfield  {author} {\bibinfo {author} {\bibfnamefont {D.}~\bibnamefont
  {Braak}},\ }\bibfield  {title} {\enquote {\bibinfo {title} {Integrability of
  the rabi model},}\ }\href {\doibase 10.1103/PhysRevLett.107.100401}
  {\bibfield  {journal} {\bibinfo  {journal} {Phys. Rev. Lett.}\ }\textbf
  {\bibinfo {volume} {107}},\ \bibinfo {pages} {100401} (\bibinfo {year}
  {2011})}\BibitemShut {NoStop}%
\bibitem [{\citenamefont {Dmytruk}\ and\ \citenamefont
  {Schir\'o}(2021)}]{Dmytruk2021}%
  \BibitemOpen
  \bibfield  {author} {\bibinfo {author} {\bibfnamefont {Olesia}\ \bibnamefont
  {Dmytruk}}\ and\ \bibinfo {author} {\bibfnamefont {Marco}\ \bibnamefont
  {Schir\'o}},\ }\bibfield  {title} {\enquote {\bibinfo {title} {Gauge fixing
  for strongly correlated electrons coupled to quantum light},}\ }\href
  {\doibase 10.1103/PhysRevB.103.075131} {\bibfield  {journal} {\bibinfo
  {journal} {Phys. Rev. B}\ }\textbf {\bibinfo {volume} {103}},\ \bibinfo
  {pages} {075131} (\bibinfo {year} {2021})}\BibitemShut {NoStop}%
\bibitem [{\citenamefont {Dmytruk}\ and\ \citenamefont
  {Schir{\`o}}(2022)}]{Dmytruk2022}%
  \BibitemOpen
  \bibfield  {author} {\bibinfo {author} {\bibfnamefont {Olesia}\ \bibnamefont
  {Dmytruk}}\ and\ \bibinfo {author} {\bibfnamefont {Marco}\ \bibnamefont
  {Schir{\`o}}},\ }\bibfield  {title} {\enquote {\bibinfo {title} {Controlling
  topological phases of matter with quantum light},}\ }\href {\doibase
  10.1038/s42005-022-01049-0} {\bibfield  {journal} {\bibinfo  {journal}
  {Communications Physics}\ }\textbf {\bibinfo {volume} {5}},\ \bibinfo {pages}
  {271} (\bibinfo {year} {2022})}\BibitemShut {NoStop}%
\bibitem [{\citenamefont {Lenk}\ \emph {et~al.}(2022)\citenamefont {Lenk},
  \citenamefont {Li}, \citenamefont {Werner},\ and\ \citenamefont
  {Eckstein}}]{Lenk2022}%
  \BibitemOpen
  \bibfield  {author} {\bibinfo {author} {\bibfnamefont {Katharina}\
  \bibnamefont {Lenk}}, \bibinfo {author} {\bibfnamefont {Jiajun}\ \bibnamefont
  {Li}}, \bibinfo {author} {\bibfnamefont {Philipp}\ \bibnamefont {Werner}}, \
  and\ \bibinfo {author} {\bibfnamefont {Martin}\ \bibnamefont {Eckstein}},\
  }\href {\doibase 10.48550/ARXIV.2205.05559} {\enquote {\bibinfo {title}
  {Collective theory for an interacting solid in a single-mode cavity},}\ }
  (\bibinfo {year} {2022}),\ \Eprint {http://arxiv.org/abs/2205.05559}
  {arXiv:2205.05559} \BibitemShut {NoStop}%
\bibitem [{\citenamefont {Tokatly}(2013)}]{Tokatly2013PRL}%
  \BibitemOpen
  \bibfield  {author} {\bibinfo {author} {\bibfnamefont {I.~V.}\ \bibnamefont
  {Tokatly}},\ }\bibfield  {title} {\enquote {\bibinfo {title} {Time-dependent
  density functional theory for many-electron systems interacting with cavity
  photons},}\ }\href@noop {} {\bibfield  {journal} {\bibinfo  {journal} {Phys.
  Rev. Lett.}\ }\textbf {\bibinfo {volume} {110}},\ \bibinfo {pages} {233001}
  (\bibinfo {year} {2013})}\BibitemShut {NoStop}%
\bibitem [{\citenamefont {Ruggenthaler}\ \emph {et~al.}(2014)\citenamefont
  {Ruggenthaler}, \citenamefont {Flick}, \citenamefont {Pellegrini},
  \citenamefont {Appel}, \citenamefont {Tokatly},\ and\ \citenamefont
  {Rubio}}]{Ruggenthaler2014PRA}%
  \BibitemOpen
  \bibfield  {author} {\bibinfo {author} {\bibfnamefont {Michael}\ \bibnamefont
  {Ruggenthaler}}, \bibinfo {author} {\bibfnamefont {Johannes}\ \bibnamefont
  {Flick}}, \bibinfo {author} {\bibfnamefont {Camilla}\ \bibnamefont
  {Pellegrini}}, \bibinfo {author} {\bibfnamefont {Heiko}\ \bibnamefont
  {Appel}}, \bibinfo {author} {\bibfnamefont {Ilya~V.}\ \bibnamefont
  {Tokatly}}, \ and\ \bibinfo {author} {\bibfnamefont {Angel}\ \bibnamefont
  {Rubio}},\ }\bibfield  {title} {\enquote {\bibinfo {title}
  {Quantum-electrodynamical density-functional theory: Bridging quantum optics
  and electronic-structure theory},}\ }\href@noop {} {\bibfield  {journal}
  {\bibinfo  {journal} {Phys. Rev. A}\ }\textbf {\bibinfo {volume} {90}},\
  \bibinfo {pages} {012508} (\bibinfo {year} {2014})}\BibitemShut {NoStop}%
\bibitem [{\citenamefont {Farzanehpour}\ and\ \citenamefont
  {Tokatly}(2014)}]{FarTok2014PRB}%
  \BibitemOpen
  \bibfield  {author} {\bibinfo {author} {\bibfnamefont {M.}~\bibnamefont
  {Farzanehpour}}\ and\ \bibinfo {author} {\bibfnamefont {I.~V.}\ \bibnamefont
  {Tokatly}},\ }\bibfield  {title} {\enquote {\bibinfo {title} {Quantum
  electrodynamical time-dependent density-functional theory for many-electron
  systems on a lattice},}\ }\href@noop {} {\bibfield  {journal} {\bibinfo
  {journal} {Phys. Rev. B}\ }\textbf {\bibinfo {volume} {90}},\ \bibinfo
  {pages} {195149} (\bibinfo {year} {2014})}\BibitemShut {NoStop}%
\bibitem [{\citenamefont {Pellegrini}\ \emph {et~al.}(2015)\citenamefont
  {Pellegrini}, \citenamefont {Flick}, \citenamefont {Tokatly}, \citenamefont
  {Appel},\ and\ \citenamefont {Rubio}}]{Pellegrini2015PRL}%
  \BibitemOpen
  \bibfield  {author} {\bibinfo {author} {\bibfnamefont {Camilla}\ \bibnamefont
  {Pellegrini}}, \bibinfo {author} {\bibfnamefont {Johannes}\ \bibnamefont
  {Flick}}, \bibinfo {author} {\bibfnamefont {Ilya~V.}\ \bibnamefont
  {Tokatly}}, \bibinfo {author} {\bibfnamefont {Heiko}\ \bibnamefont {Appel}},
  \ and\ \bibinfo {author} {\bibfnamefont {Angel}\ \bibnamefont {Rubio}},\
  }\bibfield  {title} {\enquote {\bibinfo {title} {Optimized effective
  potential for quantum electrodynamical time-dependent density functional
  theory},}\ }\href {\doibase 10.1103/PhysRevLett.115.093001} {\bibfield
  {journal} {\bibinfo  {journal} {Phys. Rev. Lett.}\ }\textbf {\bibinfo
  {volume} {115}},\ \bibinfo {pages} {093001} (\bibinfo {year}
  {2015})}\BibitemShut {NoStop}%
\bibitem [{\citenamefont {Flick}\ \emph {et~al.}(2015)\citenamefont {Flick},
  \citenamefont {Ruggenthaler}, \citenamefont {Appel},\ and\ \citenamefont
  {Rubio}}]{Flick2015}%
  \BibitemOpen
  \bibfield  {author} {\bibinfo {author} {\bibfnamefont {Johannes}\
  \bibnamefont {Flick}}, \bibinfo {author} {\bibfnamefont {Michael}\
  \bibnamefont {Ruggenthaler}}, \bibinfo {author} {\bibfnamefont {Heiko}\
  \bibnamefont {Appel}}, \ and\ \bibinfo {author} {\bibfnamefont {Angel}\
  \bibnamefont {Rubio}},\ }\bibfield  {title} {\enquote {\bibinfo {title}
  {Kohn-sham approach to quantum electrodynamical density-functional theory:
  Exact time-dependent effective potentials in real space},}\ }\href {\doibase
  10.1073/pnas.1518224112} {\bibfield  {journal} {\bibinfo  {journal} {PNAS}\
  }\textbf {\bibinfo {volume} {112}},\ \bibinfo {pages} {15285--15290}
  (\bibinfo {year} {2015})}\BibitemShut {NoStop}%
\bibitem [{\citenamefont {Ruggenthaler}(2017)}]{ruggenthaler2017groundstate}%
  \BibitemOpen
  \bibfield  {author} {\bibinfo {author} {\bibfnamefont {M.}~\bibnamefont
  {Ruggenthaler}},\ }\href@noop {} {\enquote {\bibinfo {title} {Ground-state
  quantum-electrodynamical density-functional theory},}\ } (\bibinfo {year}
  {2017}),\ \Eprint {http://arxiv.org/abs/1509.01417} {arXiv:1509.01417
  [quant-ph]} \BibitemShut {NoStop}%
\bibitem [{\citenamefont {Jaako}\ \emph {et~al.}(2016)\citenamefont {Jaako},
  \citenamefont {Xiang}, \citenamefont {Garcia-Ripoll},\ and\ \citenamefont
  {Rabl}}]{PhysRevA.94.033850}%
  \BibitemOpen
  \bibfield  {author} {\bibinfo {author} {\bibfnamefont {Tuomas}\ \bibnamefont
  {Jaako}}, \bibinfo {author} {\bibfnamefont {Ze-Liang}\ \bibnamefont {Xiang}},
  \bibinfo {author} {\bibfnamefont {Juan~Jos\'e}\ \bibnamefont
  {Garcia-Ripoll}}, \ and\ \bibinfo {author} {\bibfnamefont {Peter}\
  \bibnamefont {Rabl}},\ }\bibfield  {title} {\enquote {\bibinfo {title}
  {Ultrastrong-coupling phenomena beyond the dicke model},}\ }\href {\doibase
  10.1103/PhysRevA.94.033850} {\bibfield  {journal} {\bibinfo  {journal} {Phys.
  Rev. A}\ }\textbf {\bibinfo {volume} {94}},\ \bibinfo {pages} {033850}
  (\bibinfo {year} {2016})}\BibitemShut {NoStop}%
\bibitem [{\citenamefont {Pilar}\ \emph {et~al.}(2020)\citenamefont {Pilar},
  \citenamefont {De~Bernardis},\ and\ \citenamefont
  {Rabl}}]{pilar2020thermodynamics}%
  \BibitemOpen
  \bibfield  {author} {\bibinfo {author} {\bibfnamefont {Philipp}\ \bibnamefont
  {Pilar}}, \bibinfo {author} {\bibfnamefont {Daniele}\ \bibnamefont
  {De~Bernardis}}, \ and\ \bibinfo {author} {\bibfnamefont {Peter}\
  \bibnamefont {Rabl}},\ }\bibfield  {title} {\enquote {\bibinfo {title}
  {Thermodynamics of ultrastrongly coupled light-matter systems},}\ }\href@noop
  {} {\bibfield  {journal} {\bibinfo  {journal} {Quantum}\ }\textbf {\bibinfo
  {volume} {4}},\ \bibinfo {pages} {335} (\bibinfo {year} {2020})}\BibitemShut
  {NoStop}%
\bibitem [{\citenamefont {De~Bernardis}\ \emph
  {et~al.}(2018{\natexlab{a}})\citenamefont {De~Bernardis}, \citenamefont
  {Jaako},\ and\ \citenamefont {Rabl}}]{PhysRevA.97.043820}%
  \BibitemOpen
  \bibfield  {author} {\bibinfo {author} {\bibfnamefont {Daniele}\ \bibnamefont
  {De~Bernardis}}, \bibinfo {author} {\bibfnamefont {Tuomas}\ \bibnamefont
  {Jaako}}, \ and\ \bibinfo {author} {\bibfnamefont {Peter}\ \bibnamefont
  {Rabl}},\ }\bibfield  {title} {\enquote {\bibinfo {title} {Cavity quantum
  electrodynamics in the nonperturbative regime},}\ }\href {\doibase
  10.1103/PhysRevA.97.043820} {\bibfield  {journal} {\bibinfo  {journal} {Phys.
  Rev. A}\ }\textbf {\bibinfo {volume} {97}},\ \bibinfo {pages} {043820}
  (\bibinfo {year} {2018}{\natexlab{a}})}\BibitemShut {NoStop}%
\bibitem [{\citenamefont {Schuler}\ \emph {et~al.}(2020)\citenamefont
  {Schuler}, \citenamefont {Bernardis}, \citenamefont {Läuchli},\ and\
  \citenamefont {Rabl}}]{10.21468/SciPostPhys.9.5.066}%
  \BibitemOpen
  \bibfield  {author} {\bibinfo {author} {\bibfnamefont {Michael}\ \bibnamefont
  {Schuler}}, \bibinfo {author} {\bibfnamefont {Daniele~De}\ \bibnamefont
  {Bernardis}}, \bibinfo {author} {\bibfnamefont {Andreas~M.}\ \bibnamefont
  {Läuchli}}, \ and\ \bibinfo {author} {\bibfnamefont {Peter}\ \bibnamefont
  {Rabl}},\ }\bibfield  {title} {\enquote {\bibinfo {title} {{The vacua of
  dipolar cavity quantum electrodynamics}},}\ }\href {\doibase
  10.21468/SciPostPhys.9.5.066} {\bibfield  {journal} {\bibinfo  {journal}
  {SciPost Phys.}\ }\textbf {\bibinfo {volume} {9}},\ \bibinfo {pages} {066}
  (\bibinfo {year} {2020})}\BibitemShut {NoStop}%
\bibitem [{\citenamefont {Lamata}(2017)}]{lamata2017digital}%
  \BibitemOpen
  \bibfield  {author} {\bibinfo {author} {\bibfnamefont {Lucas}\ \bibnamefont
  {Lamata}},\ }\bibfield  {title} {\enquote {\bibinfo {title} {Digital-analog
  quantum simulation of generalized dicke models with superconducting
  circuits},}\ }\href@noop {} {\bibfield  {journal} {\bibinfo  {journal}
  {Scientific Reports}\ }\textbf {\bibinfo {volume} {7}},\ \bibinfo {pages}
  {1--12} (\bibinfo {year} {2017})}\BibitemShut {NoStop}%
\bibitem [{\citenamefont {Shapiro}\ \emph {et~al.}(2020)\citenamefont
  {Shapiro}, \citenamefont {Pogosov},\ and\ \citenamefont
  {Lozovik}}]{shapiro2020universal}%
  \BibitemOpen
  \bibfield  {author} {\bibinfo {author} {\bibfnamefont {DS}~\bibnamefont
  {Shapiro}}, \bibinfo {author} {\bibfnamefont {WV}~\bibnamefont {Pogosov}}, \
  and\ \bibinfo {author} {\bibfnamefont {Yu~E}\ \bibnamefont {Lozovik}},\
  }\bibfield  {title} {\enquote {\bibinfo {title} {Universal fluctuations and
  squeezing in a generalized dicke model near the superradiant phase
  transition},}\ }\href@noop {} {\bibfield  {journal} {\bibinfo  {journal}
  {Physical Review A}\ }\textbf {\bibinfo {volume} {102}},\ \bibinfo {pages}
  {023703} (\bibinfo {year} {2020})}\BibitemShut {NoStop}%
\bibitem [{\citenamefont {Gambetta}\ \emph {et~al.}(2019)\citenamefont
  {Gambetta}, \citenamefont {Lesanovsky},\ and\ \citenamefont
  {Li}}]{PhysRevA.100.022513}%
  \BibitemOpen
  \bibfield  {author} {\bibinfo {author} {\bibfnamefont {F.~M.}\ \bibnamefont
  {Gambetta}}, \bibinfo {author} {\bibfnamefont {I.}~\bibnamefont
  {Lesanovsky}}, \ and\ \bibinfo {author} {\bibfnamefont {W.}~\bibnamefont
  {Li}},\ }\bibfield  {title} {\enquote {\bibinfo {title} {Exploring
  nonequilibrium phases of the generalized dicke model with a trapped
  rydberg-ion quantum simulator},}\ }\href {\doibase
  10.1103/PhysRevA.100.022513} {\bibfield  {journal} {\bibinfo  {journal}
  {Phys. Rev. A}\ }\textbf {\bibinfo {volume} {100}},\ \bibinfo {pages}
  {022513} (\bibinfo {year} {2019})}\BibitemShut {NoStop}%
\bibitem [{\citenamefont {Akbari}\ \emph {et~al.}(2023)\citenamefont {Akbari},
  \citenamefont {Salmon}, \citenamefont {Nori},\ and\ \citenamefont
  {Hughes}}]{akbari2023generalized}%
  \BibitemOpen
  \bibfield  {author} {\bibinfo {author} {\bibfnamefont {Kamran}\ \bibnamefont
  {Akbari}}, \bibinfo {author} {\bibfnamefont {Will}\ \bibnamefont {Salmon}},
  \bibinfo {author} {\bibfnamefont {Franco}\ \bibnamefont {Nori}}, \ and\
  \bibinfo {author} {\bibfnamefont {Stephen}\ \bibnamefont {Hughes}},\
  }\bibfield  {title} {\enquote {\bibinfo {title} {Generalized dicke model and
  gauge-invariant master equations for two atoms in ultrastrongly-coupled
  cavity quantum electrodynamics},}\ }\href@noop {} {\bibfield  {journal}
  {\bibinfo  {journal} {arXiv preprint arXiv:2301.02127}\ } (\bibinfo {year}
  {2023})}\BibitemShut {NoStop}%
\bibitem [{\citenamefont {De~Bernardis}\ \emph
  {et~al.}(2018{\natexlab{b}})\citenamefont {De~Bernardis}, \citenamefont
  {Pilar}, \citenamefont {Jaako}, \citenamefont {De~Liberato},\ and\
  \citenamefont {Rabl}}]{de2018breakdown}%
  \BibitemOpen
  \bibfield  {author} {\bibinfo {author} {\bibfnamefont {Daniele}\ \bibnamefont
  {De~Bernardis}}, \bibinfo {author} {\bibfnamefont {Philipp}\ \bibnamefont
  {Pilar}}, \bibinfo {author} {\bibfnamefont {Tuomas}\ \bibnamefont {Jaako}},
  \bibinfo {author} {\bibfnamefont {Simone}\ \bibnamefont {De~Liberato}}, \
  and\ \bibinfo {author} {\bibfnamefont {Peter}\ \bibnamefont {Rabl}},\
  }\bibfield  {title} {\enquote {\bibinfo {title} {Breakdown of gauge
  invariance in ultrastrong-coupling cavity qed},}\ }\href@noop {} {\bibfield
  {journal} {\bibinfo  {journal} {Physical Review A}\ }\textbf {\bibinfo
  {volume} {98}},\ \bibinfo {pages} {053819} (\bibinfo {year}
  {2018}{\natexlab{b}})}\BibitemShut {NoStop}%
\bibitem [{\citenamefont {Andolina}\ \emph {et~al.}(2019)\citenamefont
  {Andolina}, \citenamefont {Pellegrino}, \citenamefont {Giovannetti},
  \citenamefont {MacDonald},\ and\ \citenamefont
  {Polini}}]{PhysRevB.100.121109}%
  \BibitemOpen
  \bibfield  {author} {\bibinfo {author} {\bibfnamefont {G.~M.}\ \bibnamefont
  {Andolina}}, \bibinfo {author} {\bibfnamefont {F.~M.~D.}\ \bibnamefont
  {Pellegrino}}, \bibinfo {author} {\bibfnamefont {V.}~\bibnamefont
  {Giovannetti}}, \bibinfo {author} {\bibfnamefont {A.~H.}\ \bibnamefont
  {MacDonald}}, \ and\ \bibinfo {author} {\bibfnamefont {M.}~\bibnamefont
  {Polini}},\ }\bibfield  {title} {\enquote {\bibinfo {title} {Cavity quantum
  electrodynamics of strongly correlated electron systems: A no-go theorem for
  photon condensation},}\ }\href {\doibase 10.1103/PhysRevB.100.121109}
  {\bibfield  {journal} {\bibinfo  {journal} {Phys. Rev. B}\ }\textbf {\bibinfo
  {volume} {100}},\ \bibinfo {pages} {121109} (\bibinfo {year}
  {2019})}\BibitemShut {NoStop}%
\bibitem [{\citenamefont {Andolina}\ \emph {et~al.}(2020)\citenamefont
  {Andolina}, \citenamefont {Pellegrino}, \citenamefont {Giovannetti},
  \citenamefont {MacDonald},\ and\ \citenamefont
  {Polini}}]{PhysRevB.102.125137}%
  \BibitemOpen
  \bibfield  {author} {\bibinfo {author} {\bibfnamefont {G.~M.}\ \bibnamefont
  {Andolina}}, \bibinfo {author} {\bibfnamefont {F.~M.~D.}\ \bibnamefont
  {Pellegrino}}, \bibinfo {author} {\bibfnamefont {V.}~\bibnamefont
  {Giovannetti}}, \bibinfo {author} {\bibfnamefont {A.~H.}\ \bibnamefont
  {MacDonald}}, \ and\ \bibinfo {author} {\bibfnamefont {M.}~\bibnamefont
  {Polini}},\ }\bibfield  {title} {\enquote {\bibinfo {title} {Theory of photon
  condensation in a spatially varying electromagnetic field},}\ }\href
  {\doibase 10.1103/PhysRevB.102.125137} {\bibfield  {journal} {\bibinfo
  {journal} {Phys. Rev. B}\ }\textbf {\bibinfo {volume} {102}},\ \bibinfo
  {pages} {125137} (\bibinfo {year} {2020})}\BibitemShut {NoStop}%
\bibitem [{\citenamefont {Flick}\ \emph {et~al.}(2017)\citenamefont {Flick},
  \citenamefont {Appel}, \citenamefont {Ruggenthaler},\ and\ \citenamefont
  {Rubio}}]{Flick2017b}%
  \BibitemOpen
  \bibfield  {author} {\bibinfo {author} {\bibfnamefont {Johannes}\
  \bibnamefont {Flick}}, \bibinfo {author} {\bibfnamefont {Heiko}\ \bibnamefont
  {Appel}}, \bibinfo {author} {\bibfnamefont {Michael}\ \bibnamefont
  {Ruggenthaler}}, \ and\ \bibinfo {author} {\bibfnamefont {Angel}\
  \bibnamefont {Rubio}},\ }\bibfield  {title} {\enquote {\bibinfo {title}
  {Cavity born oppenheimer approximation for correlated electron nuclear-photon
  systems},}\ }\href {\doibase 10.1021/acs.jctc.6b01126} {\bibfield  {journal}
  {\bibinfo  {journal} {J. Chem. Theory Comput.}\ }\textbf {\bibinfo {volume}
  {13}},\ \bibinfo {pages} {1616--1625} (\bibinfo {year} {2017})},\ \bibinfo
  {note} {pMID: 28277664}\BibitemShut {NoStop}%
\bibitem [{\citenamefont {Liu}\ \emph {et~al.}(2014)\citenamefont {Liu},
  \citenamefont {Petersson}, \citenamefont {Stehlik}, \citenamefont {Taylor},\
  and\ \citenamefont {Petta}}]{liu2014photon}%
  \BibitemOpen
  \bibfield  {author} {\bibinfo {author} {\bibfnamefont {Y-Y}\ \bibnamefont
  {Liu}}, \bibinfo {author} {\bibfnamefont {KD}~\bibnamefont {Petersson}},
  \bibinfo {author} {\bibfnamefont {J}~\bibnamefont {Stehlik}}, \bibinfo
  {author} {\bibfnamefont {Jacob~M}\ \bibnamefont {Taylor}}, \ and\ \bibinfo
  {author} {\bibfnamefont {Jason~R}\ \bibnamefont {Petta}},\ }\bibfield
  {title} {\enquote {\bibinfo {title} {Photon emission from a cavity-coupled
  double quantum dot},}\ }\href@noop {} {\bibfield  {journal} {\bibinfo
  {journal} {Physical review letters}\ }\textbf {\bibinfo {volume} {113}},\
  \bibinfo {pages} {036801} (\bibinfo {year} {2014})}\BibitemShut {NoStop}%
\bibitem [{\citenamefont {Deng}\ \emph {et~al.}(2015)\citenamefont {Deng},
  \citenamefont {Wei}, \citenamefont {Li}, \citenamefont {Johansson},
  \citenamefont {Kong}, \citenamefont {Li}, \citenamefont {Cao}, \citenamefont
  {Xiao}, \citenamefont {Guo}, \citenamefont {Nori} \emph
  {et~al.}}]{deng2015coupling}%
  \BibitemOpen
  \bibfield  {author} {\bibinfo {author} {\bibfnamefont {Guang-Wei}\
  \bibnamefont {Deng}}, \bibinfo {author} {\bibfnamefont {Da}~\bibnamefont
  {Wei}}, \bibinfo {author} {\bibfnamefont {Shu-Xiao}\ \bibnamefont {Li}},
  \bibinfo {author} {\bibfnamefont {JR}~\bibnamefont {Johansson}}, \bibinfo
  {author} {\bibfnamefont {Wei-Cheng}\ \bibnamefont {Kong}}, \bibinfo {author}
  {\bibfnamefont {Hai-Ou}\ \bibnamefont {Li}}, \bibinfo {author} {\bibfnamefont
  {Gang}\ \bibnamefont {Cao}}, \bibinfo {author} {\bibfnamefont {Ming}\
  \bibnamefont {Xiao}}, \bibinfo {author} {\bibfnamefont {Guang-Can}\
  \bibnamefont {Guo}}, \bibinfo {author} {\bibfnamefont {Franco}\ \bibnamefont
  {Nori}},  \emph {et~al.},\ }\bibfield  {title} {\enquote {\bibinfo {title}
  {Coupling two distant double quantum dots with a microwave resonator},}\
  }\href@noop {} {\bibfield  {journal} {\bibinfo  {journal} {Nano letters}\
  }\textbf {\bibinfo {volume} {15}},\ \bibinfo {pages} {6620--6625} (\bibinfo
  {year} {2015})}\BibitemShut {NoStop}%
\bibitem [{\citenamefont {Liu}\ \emph {et~al.}(2015)\citenamefont {Liu},
  \citenamefont {Stehlik}, \citenamefont {Eichler}, \citenamefont {Gullans},
  \citenamefont {Taylor},\ and\ \citenamefont {Petta}}]{liu2015semiconductor}%
  \BibitemOpen
  \bibfield  {author} {\bibinfo {author} {\bibfnamefont {Y-Y}\ \bibnamefont
  {Liu}}, \bibinfo {author} {\bibfnamefont {J}~\bibnamefont {Stehlik}},
  \bibinfo {author} {\bibfnamefont {Christopher}\ \bibnamefont {Eichler}},
  \bibinfo {author} {\bibfnamefont {MJ}~\bibnamefont {Gullans}}, \bibinfo
  {author} {\bibfnamefont {Jacob~M}\ \bibnamefont {Taylor}}, \ and\ \bibinfo
  {author} {\bibfnamefont {JR}~\bibnamefont {Petta}},\ }\bibfield  {title}
  {\enquote {\bibinfo {title} {Semiconductor double quantum dot micromaser},}\
  }\href@noop {} {\bibfield  {journal} {\bibinfo  {journal} {Science}\ }\textbf
  {\bibinfo {volume} {347}},\ \bibinfo {pages} {285--287} (\bibinfo {year}
  {2015})}\BibitemShut {NoStop}%
\bibitem [{\citenamefont {Cirio}\ \emph {et~al.}(2016)\citenamefont {Cirio},
  \citenamefont {De~Liberato}, \citenamefont {Lambert},\ and\ \citenamefont
  {Nori}}]{PhysRevLett.116.113601}%
  \BibitemOpen
  \bibfield  {author} {\bibinfo {author} {\bibfnamefont {Mauro}\ \bibnamefont
  {Cirio}}, \bibinfo {author} {\bibfnamefont {Simone}\ \bibnamefont
  {De~Liberato}}, \bibinfo {author} {\bibfnamefont {Neill}\ \bibnamefont
  {Lambert}}, \ and\ \bibinfo {author} {\bibfnamefont {Franco}\ \bibnamefont
  {Nori}},\ }\bibfield  {title} {\enquote {\bibinfo {title} {Ground state
  electroluminescence},}\ }\href {\doibase 10.1103/PhysRevLett.116.113601}
  {\bibfield  {journal} {\bibinfo  {journal} {Phys. Rev. Lett.}\ }\textbf
  {\bibinfo {volume} {116}},\ \bibinfo {pages} {113601} (\bibinfo {year}
  {2016})}\BibitemShut {NoStop}%
\end{thebibliography}%

\begin{widetext}

\newpage

\section*{Supplementary Material}

\subsection*{Dimer behaviour within different external field scales}

The energy and average value of the $z$-projection of the effectvie  spin are the primary observables that we are interested in. In the main text we introduced the variables $d_{(1,2)}=1/(N_{(1,2)})\sum_{i=1}^{N_{(1,2)}}\langle \sigma_{z}^{(1,2),i} \rangle$, and of the total system polarization $P=\langle S_z\rangle$. In Fig.~\ref{fig:3x3_full}, we illustrate the behavior of $d_1$ and $d_2$ in the ground state, depending on the external potential $v_{\text{ext}}$ acting on the second group of dimers~($N_2$). In the figure, we present nine combinations of dimer numbers within each group. We observe an interesting feature in the behavior of the system's polarization $P$ as a function of the external potential $v_{\text{ext}}$ acting on the second group of dimers. For small values of $v_{\text{ext}}$, $P$ takes on only integer values in steps of two. This is because we have chosen $\hat{\sigma}$ as the spin operator, which discards the contribution of $1/2$ to the polarization. We remark that this effect can only be observed when the external potential is small in magnitude. The situation on a large scale of $v_{\text{ext}}$ is shown in Fig.~\ref{fig:3x3_big_fields}. It should be noted that at high values of the external potential, the correct asymptotic behavior of the dimers can only be observed in the case of exact diagonalization.
\begin{figure}[h]
    \centering
    \includegraphics[width = 0.99\linewidth]{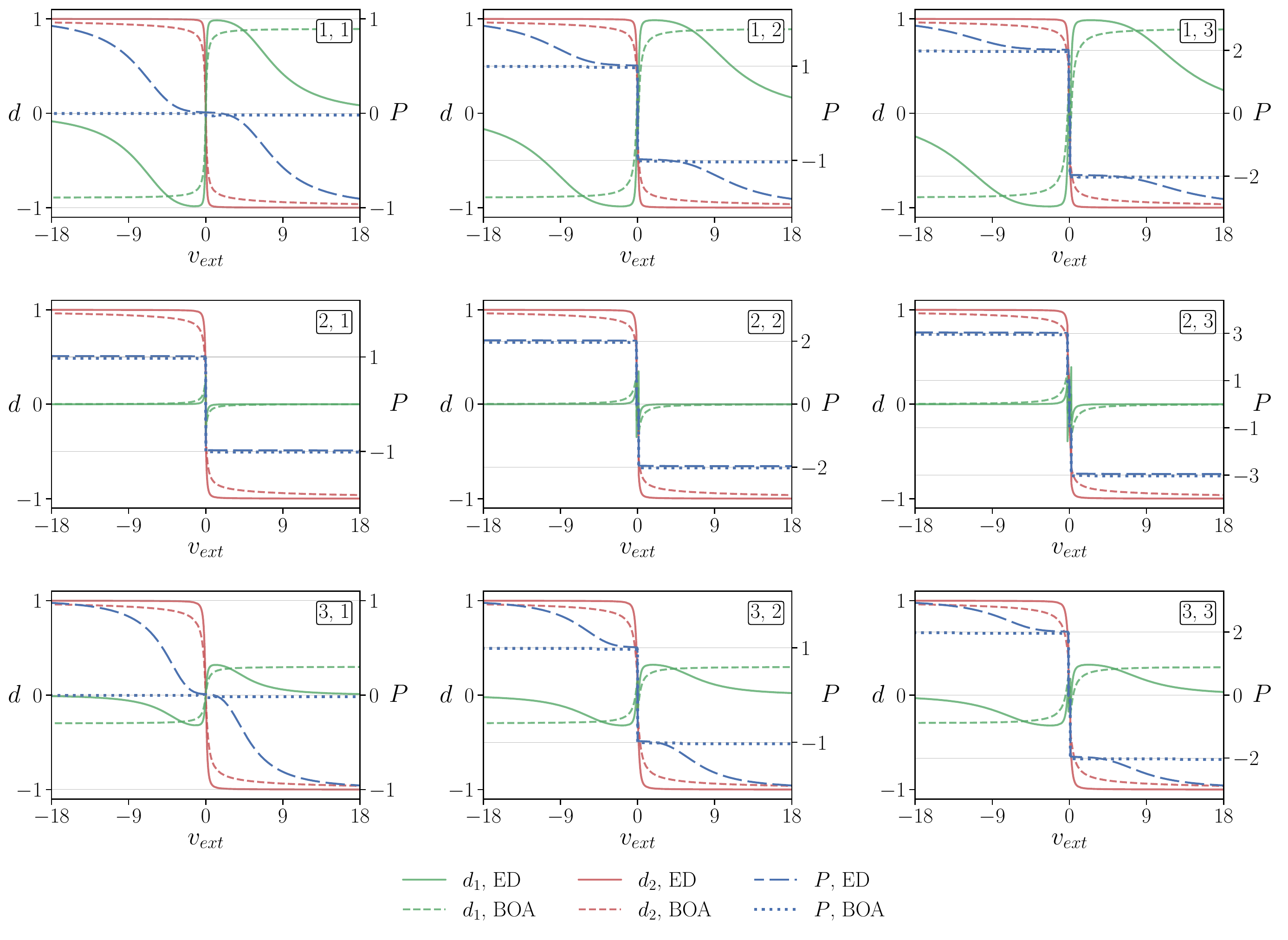}
    \caption{Dependencies of value of the $z$-projection on the external field $v_\text{ext}$ of the second group of dimers. All figures correspond to different combinations of irradiated (second group of dimers, $N_2$) and non-irradiated (first group of dimers, $N_1$) by external potential $v_{\textup{ext}}$. System parameters: $\lambda = 3$, $\omega = 1$, $T = 1$. }
    \label{fig:3x3_big_fields}
\end{figure}
\begin{figure}
    \centering
    \includegraphics[width = 1\linewidth]{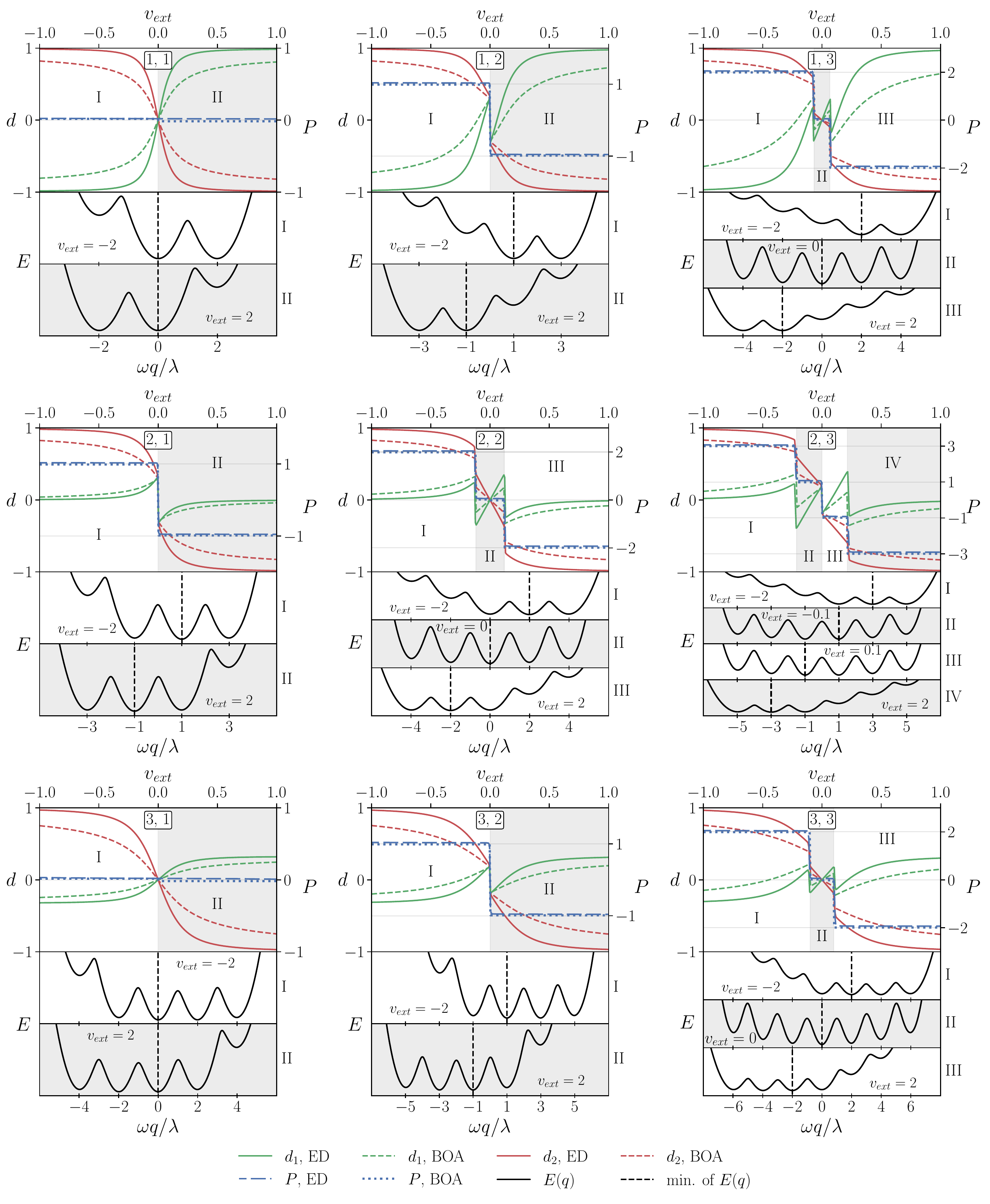}
    \caption{Dependencies of value of the $z$-projection of the spin of specific dimer within first and second groups (upper panel) on the external field $v_\text{ext}$ applied to the second group of dimers and the semiclassically calculated  system energy $E(q)$ (lower panel) on the value of photon coordinate $q$ treated in this case as $c$-number. All figures correspond to different combinations of non-irradiated (first group of dimers, $N_1$, first number in the box) and irradiated (second group of dimers, $N_2$, second number in the box)  by external potential $v_{\textup{ext}}$. The convergence of the results was achieved by cutting off the Fock space for photons by $\sim 150$. System parameters: $\lambda = 3$, $\omega = 1$, $T = 1$. }
    \label{fig:3x3_full}
\end{figure}

\section*{Polarization behaviour at different numbers of dimers}

In this section, we would like to demonstrate how the polarization of the system behaves with an increase in the total number of dimers. Namely, that the thermodynamic limit can be described using the RPA, in which the stepwise dependence of the polarization on the external potential becomes linear. To do this, in Fig.~\ref{fig:10x10_full} we present the behavior of the normalized polarization ($P/(N_1+N_2)$) depending on the external potential $v_{\textup{ext}}$ for different values of the couples $(N_1,N_2)$.

\begin{figure}
    \centering
    \includegraphics[width = 0.65\linewidth]{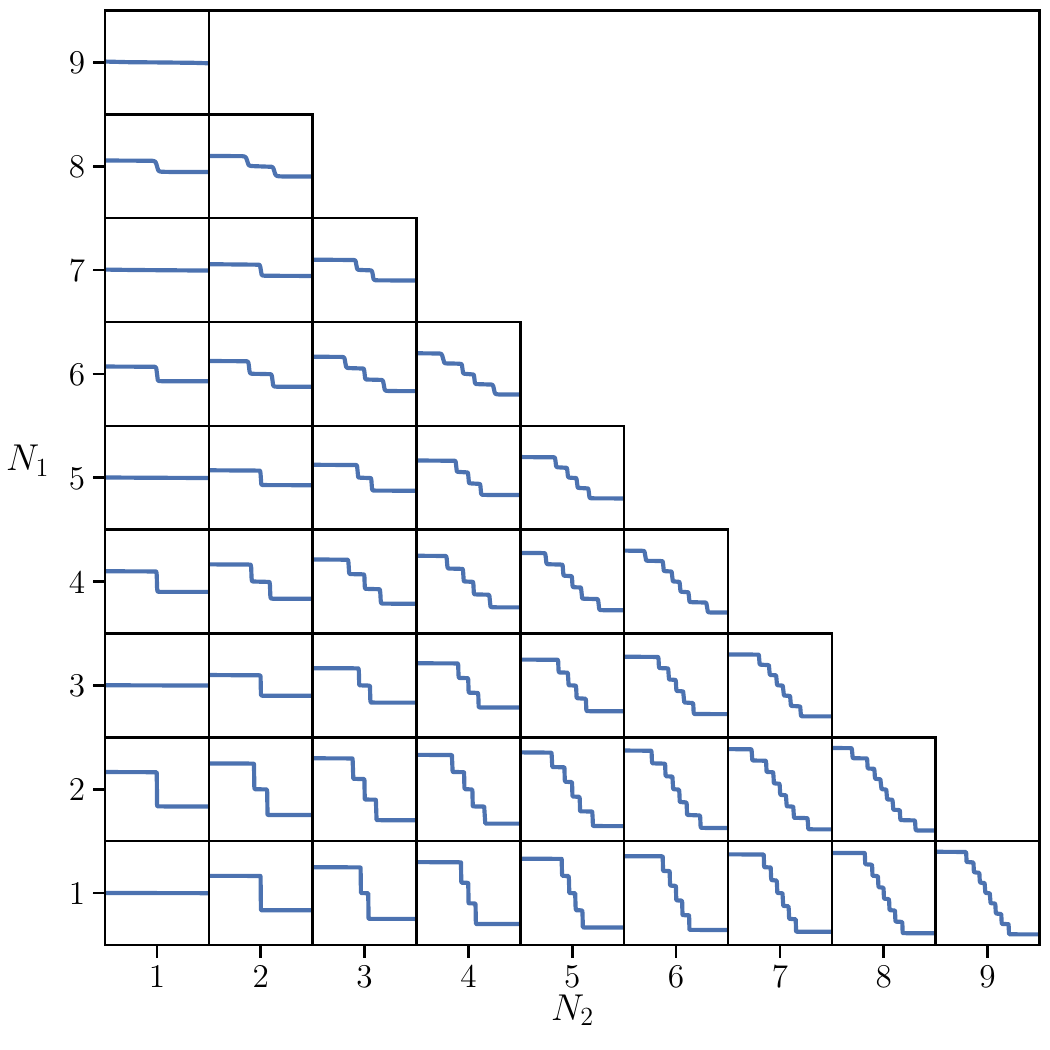}
    \caption{Dependencies of the normalized polarization (vertical axis within each subplot) on the external field $v_\text{ext}$ (horizontal axis within each subplot) applied to the second group of dimers. All insets correspond to  the different couples of dimer numbers $(N_1,N_2)$. The normalization is carried out by dividing by the total number of dimers in the system: $P/(N_1+N_2)$. The external potential $v_{\textup{ext}}$ for each subplot varies within the range $[-1,1]$. System parameters: $\lambda = 3$, $\omega = 1$, $T = 1$. }
    \label{fig:10x10_full}
\end{figure}

\section*{Two examples of diagrams}

In this section, we look at the behavior of the two diagrams mentioned in the paper. To take the thermodynamic limit, we make a variable change $N_1 \rightarrow \alpha N$, $N_2 \rightarrow \beta  N$, and $\lambda \rightarrow \lambda_0/\sqrt{N}$, with the ratio $\alpha/\beta$ is assumed to remain finite. The diagrams read as:
\begin{align}
    \vcenter{\hbox{\begin{tikzpicture}[thick, use Hobby shortcut, scale=0.36]
\draw (-1.8,-1.7) .. (-2.6,0.0) .. (-1.8,1.7);
\draw (-1.8,-1.7) .. (-1.,0.0) .. (-1.8,1.7);
  \path [draw=black,snake it]
    (-1.8,-1.7) -- (1.8,-1.7);
      \path [draw=black,snake it]
    (-1.8,1.7) -- (1.8,1.7);
\draw (1.8,-1.7) .. (2.6,0.0) .. (1.8,1.7);
\draw (1.8,-1.7) .. (1.,0.0) .. (1.8,1.7);
 \fill (1.8,-1.7) circle (1pt);
 \fill (1.8,1.7) circle (1pt);
 \end{tikzpicture}}}
&=\dfrac{N_1N_2}{(2\pi)^3}\int\limits_{-\infty}^{\infty}\sum\limits_{i=1}^2\sum\limits_{k=1}^2\sum\limits_{l=1}^2\sum\limits_{j=1}^2\dfrac{ d\varepsilon_1d\varepsilon_2 d\omega' d_{i,k}^1 d_{k,i}^1 d_{j,l}^2d_{l,j}^2 D_{\textup{ph}}^2(\omega,\omega')}{(i\varepsilon_1-E_{i}^1)(i\varepsilon_1-i\omega'-E_k^1)(i\varepsilon_2-E_j^2)(i\varepsilon_2+i\omega'-E_l^2)}\nonumber\\
&=\alpha\beta\lambda_0^4\left[\dfrac{4}{27}-\dfrac{4 v_{\textup{ext,2}}^2}{27}  -\dfrac{4 v_{\textup{ext,2}}^2}{27}  +\dfrac{38 v_{\textup{ext,1}}^2 v_{\textup{ext,2}}^2}{243}+O(v_{\textup{ext}}^4) \right],\\
\vcenter{\hbox{\begin{tikzpicture}[thick, use Hobby shortcut, scale=0.36]
\draw (-1.8,-1.7) .. (-2.6,0.0) .. (-1.8,1.7);
\draw (-1.8,-1.7) .. (-1.,0.0) .. (-1.8,1.7);
\path [draw=black,snake it]
    (-1.8,-1.7) -- (1.8,-1.7);
\path [draw=black,snake it]
    (-1.8,1.7) -- (1.8,1.7);
\path [draw=black,snake it]
    (-1,0) -- (1,0);
\draw (1.8,-1.7) .. (2.6,0.0) .. (1.8,1.7);
\draw (1.8,-1.7) .. (1.,0.0) .. (1.8,1.7);
 \fill (1.8,-1.7) circle (1pt);
 \fill (1.8,1.7) circle (1pt);
 \end{tikzpicture}}}
&=\dfrac{N_1N_2}{(2\pi)^4}\int\limits_{-\infty}^{\infty}\sum\limits_{i=1}^2\sum\limits_{k=1}^2\sum\limits_{j=1}^2\sum\limits_{o=1}^2\sum\limits_{m=1}^2\sum\limits_{n=1}^2\nonumber\\
&\qquad\qquad\times\dfrac{d\varepsilon_1d\varepsilon_2 d\omega'd\omega''d_{i,k}^1 d_{k,j}^1 d_{j,i}^1 d_{m,o}^2 d_{o,n}^2 d_{n,m}^2D_{\textup{ph}}(\omega,\omega')D_{\textup{ph}}(\omega,\omega'-\omega'')D_{\textup{ph}}(\omega,\omega'')}{(i\varepsilon_1-E_{i}^1)(i\varepsilon_1-i\omega'-E_j^1)(i\varepsilon_1-i\omega''-E_k^1)(i\varepsilon_2-E_m^2)(i\varepsilon_2-i\omega'-E_n^2)(i\varepsilon_2-i\omega''-E_o^2)}\nonumber\\
&=\alpha\beta\dfrac{\lambda_0^6}{N}\left[-\dfrac{38 v_{\textup{ext,1}} v_{\textup{ext,2}}}{30}+\dfrac{23 v_{\textup{ext,1}}^3 v_{\textup{ext,2}}}{450}+\dfrac{23 v_{\textup{ext,1}} v_{\textup{ext,2}}^3}{450}+
+O(v_{\textup{ext}}^6) \right],
\end{align}
where photon propagator, dipole matrix elements, and energies are defined as:
\begin{align}
&D_{\textup{ph}}(\omega,\omega')=-\dfrac{\lambda^2\omega'^2}{\omega'^2+\omega^2},\quad  d_{1,2}^1=d_{2,1}^1=\dfrac{T}{W(v_{\textup{ext},1}, T)}, \quad  d_{1,2}^2=d_{2,1}^2=\dfrac{T}{W(v_{\textup{ext},2}, T)}, \quad  W(v_{\textup{ext}}, T)=\sqrt{v_{\textup{ext}}^2+T^2} \nonumber \\ & d_{1,1}^1=-\dfrac{v_{\textup{ext},1}}{W(v_{\textup{ext},1}, T)}, \quad
d_{2,2}^1=\dfrac{v_{\textup{ext},1}}{W(v_{\textup{ext},1}, T)}, \quad  d_{1,1}^2=-\dfrac{v_{\textup{ext},2}}{W(v_{\textup{ext},2}, T)}, \quad
d_{2,2}^2=\dfrac{v_{\textup{ext},2}}{W(v_{\textup{ext},2}, T)},\nonumber\\
&E_{1}^1 = - W(v_{\textup{ext},1},T), \quad E_{2}^1 = W(v_{\textup{ext},1},T), \quad E_{1}^2 = - W(v_{\textup{ext},2},T), \quad E_{2}^2 = W(v_{\textup{ext},2},T).
\end{align}
\end{widetext}

\end{document}